\documentclass[letterpaper,twocolumn,10pt]{article}
\usepackage{usenix}

\usepackage{tikz}
\usepackage{amsmath}
\usepackage{color}
\usepackage{booktabs}
\usepackage{paralist}
\usepackage{multirow}
\usepackage{longtable}
\usepackage{graphicx}
\usepackage{soul}
\usepackage{footnote}
\usepackage[ruled,linesnumbered]{algorithm2e}

\usepackage{filecontents}
\begin{document}

\date{}

\title{\Large \bf BackdoorIndicator: Leveraging OOD Data for \\ Proactive
Backdoor Detection in Federated Learning}

\author{
\rm Songze Li\thanks{The two authors contributed equally to this work.}\\
Southeast University\\
\textit{songzeli@seu.edu.cn}
\and
\rm Yanbo Dai{\color{green!80!black}$^*$}\\
HKUST(GZ)\\
\textit{ydai851@connect.hkust-gz.edu.cn}
}

\maketitle

\begin{abstract}
In a federated learning (FL) system, decentralized data owners (clients) could
upload their locally trained models to a central server, to jointly train a
global model. Malicious clients may plant backdoors into the global model
through uploading poisoned local models, causing misclassification to a target
class when encountering attacker-defined triggers. Existing backdoor defenses
show inconsistent performance under different system and adversarial settings,
especially when the malicious updates are made statistically close to the benign
ones. In this paper, we first reveal the fact that planting subsequent backdoors
with the same target label could significantly help to maintain the accuracy of
previously planted backdoors, and then propose a novel proactive backdoor
detection mechanism for FL named BackdoorIndicator, which has the server inject
indicator tasks into the global model leveraging out-of-distribution (OOD) data,
and then utilizing the fact that any backdoor samples are OOD samples with
respect to benign samples, the server, who is completely agnostic of the
potential backdoor types and target labels, can accurately detect the presence
of backdoors in uploaded models, via evaluating the indicator tasks. We perform
systematic and extensive empirical studies to demonstrate the consistently
superior performance and practicality of BackdoorIndicator over baseline
defenses, across a wide range of system and adversarial settings.
\end{abstract}

\section{Introduction}
\label{intro}
Federated learning (FL) \cite{mcmahan2017communication} is an emerging
collaborative training paradigm, which enables the use of computing resources
from multiple data owners to jointly train a global model under the coordination
of a center server. For every global round in FL, decentralized clients locally
train models using private training data, and then provide the server with these
model updates, instead of their raw data, for aggregation to generate the global
model for the next round. Despite of respecting participants' privacy, FL systems
have shown to be vulnerable to a series of malicious attacks, especially
poisoning attacks \cite{bagdasaryan01, Zhengming01, fang2020local,
geiping2020inverting, fung2020limitations,
nasr2019comprehensive, bhagoji2019analyzing}.

Adversaries could participate in the training procedure of FL to upload
carefully designed updates to manipulate the global model's performance.
Poisoning attacks can be categorized into untargeted attacks
\cite{fang2020local, shejwalkar2021manipulating, chen2017distributed,
guerraoui2018hidden} and backdoor attacks \cite{bhagoji2019analyzing,
bagdasaryan01}. While untargeted attackers aim to compromise the overall
performance of the global model, backdoor attackers try to have the model
misclassify to a particular target label only when encountering predefined
triggers. Previous works have successfully planted backdoors in FL models. For instance, Bagdasaryan et \textit{al.} \cite{bagdasaryan01} replace the global
model with the backdoor model, through uploading a poisoned model whose parameters are
scaled up to cancel the effect of benign updates during model
aggregation; Wang et \textit{al.} \cite{Hongyi01} propose to plant edge-case backdoors utilizing data which lives in the tail of the input distribution, to enhance the backdoor's durability and stealth.
Compared with untargeted attacks, backdoor
attacks are especially destructive due to their stealthy nature, as they do not
influence the model behavior when backdoor triggers are not present. This makes
effectively defending backdoor injection a challenging problem.

Previous methods to defend against backdoor attacks in FL mainly rely on modification and scrutinisation on the received model updates, and fall into two categories: 1)
influence reduction \cite{zitengsun01, cao2021provably,
McMahan2017LearningDP,Naseri2020LocalAC, Yin2018ByzantineRobustDL} and 2)
detection and filtering \cite{rieger2022deepsight, blanchard2017machine,
nguyen2022flame, foolsgold, wang2022rflbat, munoz2019byzantine, shen2016auror,
zhao2020shielding}. Influence reduction based methods generally assume that
backdoor updates are the minority compared to benign updates. Thus, the influence of
backdoor updates on the global model can be restricted by either constraining
the norm of model updates to an agreed bound, or adding a sufficient amount of
noise to the global model. These methods can slow down the rate of the backdoor
injection. However, a strong attacker who can keep participating in the training
procedure, or can control multiple clients in a single global round can still
effectively inject the backdoor \cite{zitengsun01, Zhengming01}; Detection and filtering methods build upon the assumption that introducing backdoor tasks makes the uploaded
model different from benign updates in the parameter space. These methods focus on designing mechanisms to identify backdoor updates, based on certain
distance metrics evaluated on received model parameters
\cite{nguyen2022flame,fung2020limitations}. Several methods (e.g.,~\cite{rieger2022deepsight}) also try to detect backdoor updates through manually
defined fine-grained features. Uploaded models with abnormal values in these
features are marked as suspicious, and then ruled out from aggregation. 
\vspace{-0.15cm}
\begin{table}[htbp]
  \centering
  \vspace{-0.4cm}
  \caption{Detection performance of FLAME when the attacker adopts different poisoned learning rates (plrs). The performance is evaluated through true positive rate (TPR), false postive rate (FPR) and backdoor accuracy (BA).}
  \scalebox{0.9}{
    \begin{tabular}{c||ccc}
    \toprule
    plr & TPR & FPR & BA \\
    \midrule
    0.01  & 0.0 & 44.2 & 83.0 \\
    0.025 & 1.6 & 44.0 & 88.1 \\
    0.04  & 38.8 & 40.1 & 90.8 \\
    0.055 & 100.0 & 33.6 & 13.0 \\
    \bottomrule
    \end{tabular}
    }
  \label{tab:intro}
  \vspace{-0.4cm}
\end{table}%

Detecting and suppressing backdoors by purely examining and comparing model parameters may fall short, under certain system settings and adversarial strategies. 
For instance, for a highly non-IID data distribution, 
the poisoned update from some backdoor attack with a small learning rate can be even closer to a benign update in parameter space, compared with other benign updates, making it impossible to detect this poisoned update. We empirically demonstrate the insufficiency of the current backdoor defenses through training a CIFAR10
\cite{krizhevsky2009learning} model on an FL system with 100 clients. 
The dataset is partitioned onto the clients following a non-IID fashion, using Dirichlet sampling \cite{hsu2019measuring} with parameter $\alpha=0.2$. In each global round, 10 clients are randomly
selected to contribute to model update. Each benign client locally trains for 2 iterations
with a learning rate of $0.05$. A vanilla backdoor attacker controlling a single
client tries to inject pixel-pattern backdoor \cite{gu2020badnets} into the
global model, through locally training the model on the dataset containing backdoor samples. We evaluate the detection performance of FLAME \cite{nguyen2022flame}, 
the state-of-the-art (SOTA) FL backdoor defense method, using the metrics of true positive rate (TPR): the ratio between the number of detected backdoor updates and the
total number of backdoor updates; false positive rate (FPR): the ratio between the number of incorrectly detected benign updates and the total number of benign updates; and backdoor accuracy (BA).
As shown in Table \ref{tab:intro}, while FLAME successfully detects all backdoor updates when the attacker uses a comparable poisoned learning rate (plr) of $0.055$, the TPR decays quickly to $0$ as the attacker reduces its learning rate. This leads to a backdoor accuracy of more than $80\%$,  rendering the FLAME defense ineffective. Also, the FPRs across all settings are quite high.
We note that the case could be even worse if the attacker plants more stealthy backdoors,
e.g., edge-case backdoor \cite{Hongyi01}, 
and Chameleon \cite{chameleon}. \emph{Considering the inherent limitation of the detection frameworks based on the examination and comparison of model parameters, a new detection paradigm is needed to provide consistent defense performance across different settings of FL training.} 

In this work, we first conduct investigation on the effects of multiple
backdoors on each other when planted sequentially. We observe that planting
subsequent backdoor task with the same target label could help to maintain
the persistence of the previously planted backdoors, once the misleading effect caused by batch normalization (BN) statistics shift is eliminated. Motivated by
this observation, we propose a proactive backdoor detection method,
\textit{BackdoorIndicator}, through utilizing the intrinsic property of backdoor
tasks, which is that backdoor samples are out-of-distribution (OOD) samples with
respect to benign samples from the target class. Specifically, the server will
first prepare an indicator dataset using OOD data. At the beginning of each global round, the server injects an indicator task into the global model based on the indicator dataset, and then broadcasts the global model to selected clients. After receiving model updates from clients, the
server checks the accuracy of the indicator task after correcting the BN statistics shift. The server marks those updates, whose indicator
accuracy is above some threshold, as suspicious, and rules them out from aggregation.

We provide extensive experimental results on three image datasets EMNIST \cite{cohen2017emnist}, CIFAR10 and CIFAR100 \cite{krizhevsky2009learning}, with three model architectures VGG16
\cite{simonyan2014very}, ResNet18, and ResNet34 \cite{he2016deep}. We evaluate the performance of BackdoorIndicator under various adversarial scenarios, where the attacker could control either a single or multiple clients, upload different types of backdoors using different
training algorithms, and adopt different poisoned learning rates. 
We also perform experiments for different non-IID
degrees of the FL system. We compare with several SOTA
detection methods to demonstrate the consistent superiority of BackdoorIndicator. We further explore the influence of several key
hyper-parameters on the detection performance of the proposed method, showing
that BackdoorIndicator is easy to implement, and does not exert much computation
overhead to the FL system.

In summary, our contribution is of three folds:
1) We reveal the effect that subsequently planted backdoors could help to maintain the accuracy of previously planted backdoors, which sheds
light on a new paradigm to design backdoor detection methods;
2) We propose a novel proactive backdoor detection mechanism for FL, BackdoorIndicator, through utilizing the intrinsic property of backdoor tasks and leveraging OOD data to identify backdoor updates;
3) We provide extensive empirical results to show that BackdoorIndicator consistently outperforms five SOTA backdoor detection methods across various adversarial and system scenarios.

\vspace{-0.1cm}
\section{Background}
\label{relatedwork}

\subsection{Federated Averaging}
FedAVG \cite{bagdasaryan01} is the baseline algorithm for implementing FL systems.
Specifically, we assume that each client $i$ holds a local dataset $D_i$.
FedAVG aims to minimize the summation of the local empirical losses
$\sum_{i=1}^K\mathcal{L}_i(\theta)$ of $K$ participating clients in a
decentralized manner, where we denote $\mathcal{L}_i$ as the cross-entropy loss
over $D_i$, and $\theta$ as the global model. In each global round $t$, the
server first broadcasts the current global model $\theta^t$ to a subset $S_t$ of
randomly selected clients. Each selected client then trains the local model
$\theta_i^t$ based on $\theta^t$ through optimizing $\mathcal{L}_i$ over its
local dataset $D_i$, and then sends $\theta_i^t$ backdoor to the central server.
After receiving all updates from clients, the server then aggregates all updates
through $\theta^{t+1}=\frac{1}{|S_t|}\sum_{i \in S_t}\theta^t_i$ to get the
global model for the next round. In the rest of the paper, we use FedAVG for FL
training.

\subsection{Backdoor Training Algorithms in FL}
A backdoor attacker could choose different malicious training algorithms to
inject the backdoor model. The attacker could simply replace a part of
benign samples with constructed backdoor samples, and perform mini-batch
stochastic gradient descent on the mixed training data. 
We denote this method as the vanilla backdoor algorithm. Based on the 
vanilla method, previous works have proposed more advanced methods to escape
from defenses. The attacker could train the backdoor model using projected
gradient descent (PGD) \cite{zitengsun01}, where in each iteration the backdoor
model is trained 
and then projected onto an $\ell_2$ ball 
around the model of the previous iteration. This could help to escape from the
norm-clipping defense which regularizes the norm of each received model update
within a norm bound. The attacker could also adopt Neurotoxin~\cite{Zhengming01}
or Chameleon~\cite{chameleon} to inject more durable backdoors in FL. To
implement Neurotoxin, the attacker first computes gradients over its benign
dataset, and identifies the top-$k\%$ coordinates of benign gradients that are
frequently updated by benign samples; the attacker proceeds to exclude these
parameters from backdoor training, through projecting the backdoor model,
initially trained in an unconstrained way, onto the bottom-(1-$k\%$) parameters.
Chameleon tries to enhance the backdoor durability in FL through utilizing
sample relationships. It identifies two types of samples which dominate backdoor
persistence in FL, and utilizes contrastive learning to adjust the distance
between backdoor samples and these two types of samples in feature space.
Without modifying the model architecture, it proceeds to train the classifier of
the model while freezing the feature encoder.

Previous studies have also proposed customized backdoor attacks, for attackers who control multiple clients. Specifically, to
plant pixel-pattern backdoors in a distributed manner, DBA \cite{xie2019dba} is
designed 
to decompose the global trigger into separate local patterns, which are then
embedded into the datasets of multiple corrupted clients. 3DFed
\cite{li20233dfed} is a recently proposed backdoor attack that achieves SOTA
performance against various defense mechanisms. It proposes to camouflage the
backdoor model through backdoor training with constrained loss, noise mask, and
decoy model. The method first trains the backdoor model through adding a
constraining term, which is proportional to the Euclidean distance between the
trained backdoor model and the global model, to the loss function. It then adds
each backdoor model with some noise to hide features that can be identified by
the central server. The sum of these noise masks is zero so that the global
model will not be influenced if all backdoor updates are accepted into
aggregation. For detection mechanisms which apply dimensionality reduction
techniques, e.g., principal component analysis (PCA), 3DFed proposes to upload
extra decoy models to fool such mechanisms to select garbage dimensions on which
backdoor updates are not separated from benign ones. The hyper-parameters of
these three modules can also be dynamically adjusted through acquiring feedback
from indicators planted in previous rounds. 

\subsection{Backdoor Defenses in FL}\label{sec:defenses}
We proceed to introduce several SOTA backdoor detection mechanisms in the
following. 

\noindent \textbf{Multi-Krum} \cite{blanchard2017machine} is a byzantine-robust
aggregation protocol which is initially proposed to ensure the convergence of
distributed stochastic gradient descent, under untargeted attackers.
Specifically, for $n$ participating clients with IID data, Multi-Krum could
effectively ensure convergence with $f$ malicious clients in a single round, as
long as $2f+2<n$. For every received model update, the server first identifies
$n-f-1$ closest updates in Euclidean distance, and then summarizes them to
compute a score. The update with the lowest score is added to the candidate set.
The server then iterates to perform selection with the rest of the updates until $m$
updates are selected. The global model for the next round is computed through
aggregating the updates from the candidate set.

\noindent \textbf{Deepsight} \cite{rieger2022deepsight} attempts to identify
backdoor updates through measuring the fine-grained difference between model
updates. It generally assumes that the training data of backdoor models exhibits
less heterogeneity than that of benign models. Thus, it proposes several
metrics, including Division Differences (DDifs) and Normalized Energy Updates (NEUPs), to reveal the distribution of labels in the used training data. While DDifs measures the difference between the predicted scores
of the local and global models, NEUPs analyzes the total magnitude of the
updates for the individual neurons of the output layer. After computing these
metrics, the method divides all received updates into clusters based on DDifs,
NEUPs, and cosine similarity. Subsequently, it marks all updates with
considerably low NEUPs as poisoned, and high NEUPs as benign. The method then
accepts all updates in the cluster with sufficient amount of benign models.
Deepsight also enforces a maximal $\ell_2$ norm of all updates and clip them to
the agreed bound if necessary. Finally, the final accepted updates are
aggregated to the new global model.

\noindent \textbf{Foolsgold} \cite{fung2020limitations} assumes that the
diversity of the model updates can be utilized to separate malicious updates
from benign ones, as benign clients have unique data distributions while
attackers share the same objective. Specifically, the method maintains a history
of updates from each client, and computes cosine similarity between pair-wise
historical updates. Instead of directly ruling out suspicious updates, Foolsgold
mitigates attacks through assigning low weight for updates with large cosine
similarity with others during aggregation.

\noindent \textbf{RFLBAT} \cite{wang2022rflbat} assumes that the difference
between benign and backdoor updates can be amplified and further identified
through dimensionality reduction.
The server adopting RFLBAT filters out suspicious updates in two rounds. For the
first round, the server performs PCA on all received updates $w_i$ to get
dimension-reduced updates $w_i^{\prime}$. The server then computes pair-wise
Euclidean distance based on $w_i^{\prime}$, and filters out outliers. For the
second round, the server divides these accepted updates into clusters using
K-means \cite{kmeans}, and selects the optimal cluster based on cosine
similarity. The server proceeds to exclude outliers in the optimal cluster using
pair-wise Euclidean distance to get finally accepted updates.

\noindent \textbf{FLAME} \cite{nguyen2022flame} estimates and injects sufficient
amount of noise to eliminate backdoors. The amount of noise needed is minimized
through dynamically filtering out backdoor updates and restricting update norms.
Specifically, the method first carries out model clustering to identify and
filter out suspicious updates with large deviation. The accepted updates are
then imposed a maximal $\ell_2$ norm, and are then clipped to the median of the norm of all accepted updates $m$. Finally, the server adds a Gaussian noise ${\cal N}(0,\sigma^2)$ to each update, where $\sigma=\frac{m}{\epsilon} \cdot
\sqrt{2\ln\frac{1.25}{\delta}}$ for some privacy budgets $\epsilon$ and $\delta$.

\subsection{Motivation and Overview of Solution}
We highlight that the above FL backdoor detection and removal methods are
based on computing some statistics of the uploaded model parameters. This basic
approach of examining model statistics makes these methods inherently
ineffective, in scenarios where attackers can fabricate backdoor updates that
are statistically close or even identical to the benign ones, as demonstrated
by our motivating experiments in Introduction. As a major contribution of this
work, we propose a fundamentally different principle in designing backdoor
defenses in FL, based upon two key observations: 1) all backdoor samples are
essentially out-of-distribution (OOD) data to the benign dataset; and 2)
subsequently injected backdoors on the same target label (not necessarily the
same trigger) can help to maintain the accuracy of the previously injected
backdoors. Consequently, we propose a novel FL backdoor defense mechanism
BackdoorIndicator, which has the server train the global model on OOD data as an
indicator task before sending it to the clients, and detect for any potential
backdoor injections via evaluating the indicator task. In contrast to
statistical defenses, BackdoorIndicator remains effective across all FL settings
and adversarial strategies, as the leveraged observations hold true as long as
there is a backdoor attempt, for arbitrary data distributions across clients.


\section{BackdoorIndicator}
\label{method}

\subsection{Threat Model}
We begin to formalize the setting through defining the threat model in terms of
the goal and capability of both the attacker and the defender. 

\noindent \textbf{Attacker's Goal and Capability.} The attacker aims to inject backdoors
into the FL model by corrupting local clients, making the model misbehave when
encountering certain backdoor triggers while leaving other tasks uninfluenced.
Once the attacker successfully corrupts a client, it has full
control over the client's training and model uploading process. In this paper,
we focus on targeted backdoor attackers who aim to
make the model misclassify all images with the backdoor trigger into a specific
class, denoted as the \textit{target class}. The attacker could begin the
poisoning from any global round in a continuous fashion.
Notably, we leave it free for the attacker to choose different types of backdoors
for injection. We also do not make constraints on the
number of participating clients the attacker could corrupt in every global
round. As far as we know, our paper assumes the strongest attacker ever
among all backdoor detection works, which makes the setting much more
realistic and challenging for successful defense. \\
\textbf{Defender's Goal and Capability.} The defender intends to detect
backdoors embedded in uploaded local models through certain defense protocol, and rules out potential malicious models from aggregation. The defender does not have access to either the raw data or the data distribution of local clients, but
does have white-box access to models uploaded from participating clients. We assume that the defender has no access to data which is in the
same distribution with the raw data of local clients.

\subsection{Key Intuition}
\label{motivation}

We illustrate the key intuition behind our proposed defense, through
experimentally exploring the effects of multiple backdoors on each other, when
planted sequentially. We focus on investigating how the subsequent backdoor task
influences those who have been planted in the model.

\begin{figure}[h]
    \begin{center}
    \scalebox{0.95}{
    \centerline{\includegraphics[width=\columnwidth]{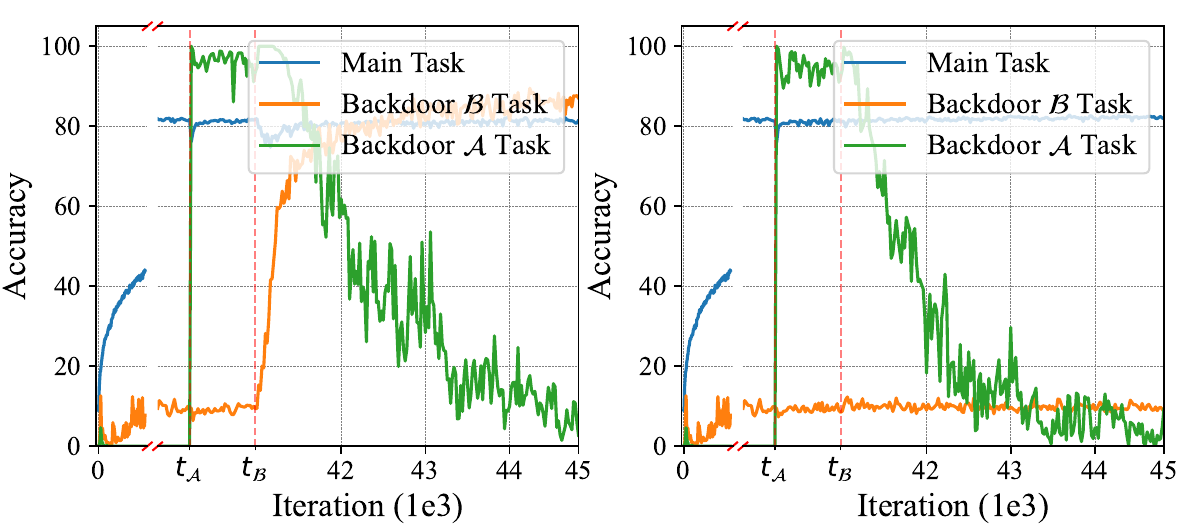}}
    }
    \caption{Accuracy of backdoor $\mathcal{A}$ task with (\textbf{Left}) and
    without (\textbf{Right}) backdoor $\mathcal{B}$ task involved after
    iteration $t_\mathcal{B}$.}
    \label{motivation_dont_replace_bn}
    \vspace{-0.7cm}
    \end{center}
\end{figure}

In the following, we consider a centralized setting where an attacker has
white-box access to the training process. Specifically, we assume that the
adversary is running an image classification task on CIFAR10 using
ResNet18. While training the main task, the adversary tries to
\textit{sequentially inject two different backdoors}, which are the
car-with-vertically-striped-walls-in-the-background semantic backdoor \cite{bagdasaryan01} as
task $\mathcal{A}$, and the pixel-pattern backdoor \cite{gu2020badnets} as task
$\mathcal{B}$, into the model. The backdoor poisoning is done by mixing up
benign images $(\boldsymbol{x}_b, y_b)$ and backdoor images $(\boldsymbol{x}_p,
y_p)$ to construct the malicious training dataset $D_1=\{\{(\boldsymbol{x}_b, y_b)^i\}_{i=1}^N, \{(\boldsymbol{x}_{p}, y_{p})^j\}_{j=1}^M\}$ with $N$ benign images and $M$ backdoor images. The attacker first injects backdoor
$\mathcal{A}$ into the model starting from iteration $t_{\mathcal{A}}$. When the
training iteration reaches $t_{\mathcal{B}}$, backdoor $\mathcal{A}$ will not be
further injected. Instead, images from backdoor $\mathcal{B}$ will be
supplemented to construct malicious training dataset $D_2=\{\{(\boldsymbol{x}_b, y_b)^i\}_{i=1}^N, \{(\boldsymbol{x}_p^\mathcal{B}, y_p^\mathcal{B})^j\}_{j=1}^M\}$. Notably, two
backdoor tasks share the same target label, i.e.,
$y_p^\mathcal{A}=y_p^\mathcal{B}$. To better illustrate the influence of
injecting backdoor $\mathcal{B}$ on the backdoor $\mathcal{A}$, we also consider
another scenario where backdoor $\mathcal{B}$ will no longer be introduced after
iteration $t_\mathcal{B}$. The variation of accuracy with training iteration
for different tasks is shown in Figure \ref{motivation_dont_replace_bn}. While
the accuracy of backdoor $\mathcal{A}$ gradually declines to 0 after the
adversary stops poisoning, planting subsequent backdoor $\mathcal{B}$ has
marginal effect in maintaining the accuracy of  $\mathcal{A}$, compared
to stopping injecting $\mathcal{B}$ after iteration $t_\mathcal{B}$. 

\begin{figure}[!h]
    \begin{center}
    \scalebox{0.95}{
    \centerline{\includegraphics[width=\columnwidth]{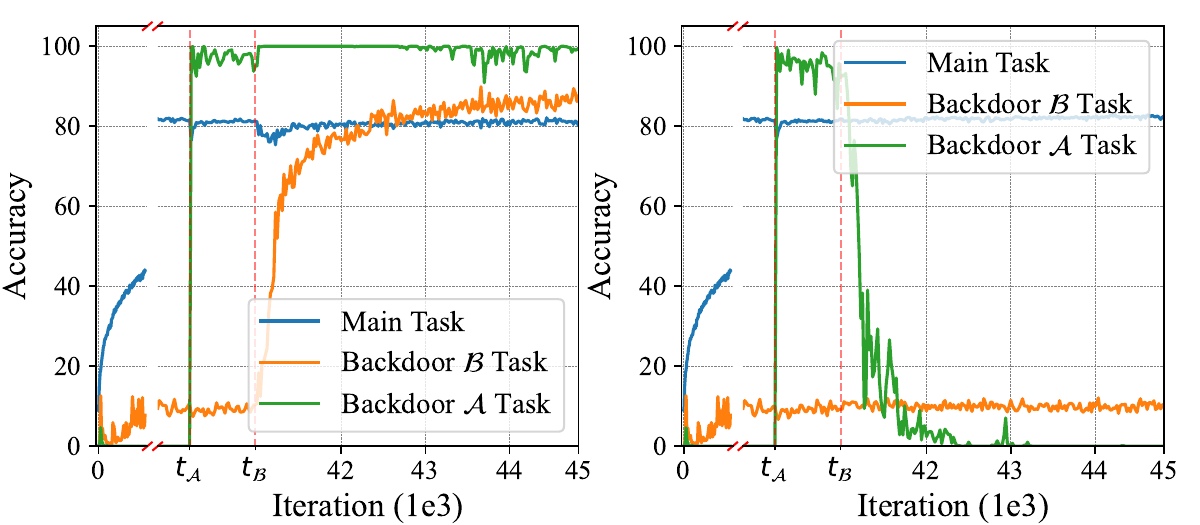}}
    }
    \vspace{-0.3cm}
    \caption{Accuracy of backdoor $\mathcal{A}$ with (\textbf{Left}) and
    without (\textbf{Right}) backdoor $\mathcal{B}$ involved after
    iteration $t_\mathcal{B}$. BN statistics are replaced with the estimated
    statistics in iteration $t_\mathcal{B}-1$.} 
    \label{motivation_replace_bn}
    \vspace{-0.4cm}
    \end{center}
\end{figure}

Nevertheless, we find that the maintaining effect shown in Figure~\ref{motivation_dont_replace_bn} is rather misleading
due to \textit{batch normalization (BN) statistics shift}. BN \cite{ioffe2015batch} is a widely employed technique to facilitate and stabilize model training. During training, for each training batch, the BN layer normalizes the output of the former layer, utilizing the empirical mean and variance computed from the batch. A running average of the mean and the variance across all batches is utilized for normalization during inference. 
In such a case, due to change of the training data, the estimated mean and variance
after iteration $t_\mathcal{B}$ gradually deviate from those of task $\mathcal{A}$, and are thus no longer applicable in evaluating
the maintaining effect. To overcome such \textit{BN statistics shift}, we save
the estimated running mean $\boldsymbol{\mu}_{t_\mathcal{B}-1}$ and variance
$\boldsymbol{\sigma}_{t_\mathcal{B}-1}$ in iteration $t_\mathcal{B}-1$, and
further replace the estimated BN statistics with
$\boldsymbol{\mu}_{t_\mathcal{B}-1}$ and $\boldsymbol{\sigma}_{t_\mathcal{B}-1}$
in evaluating task $\mathcal{A}$. As it is shown in Figure
\ref{motivation_replace_bn}, \emph{planting the subsequent backdoor task with
the same target label significantly helps to maintain the accuracy of the
previously planted backdoor.} In contrast, the accuracy of backdoor
$\mathcal{A}$ task quickly fades when there is no subsequent backdoor injected after
iteration $t_\mathcal{B}$.\footnote{Note that the observations made here are for this particular example to motivate our design, but may not hold for arbitrary backdoor combinations.}

These experimental results reveal the fact that under appropriate evaluation,
the persistence of the previously injected backdoor could be tremendously
enhanced by the subsequently planted backdoor. The reason behind the maintaining
effect touches the intrinsic property of backdoor tasks - \textbf{backdoor
samples are out-of-distribution (OOD) samples with respect to benign samples
from the target class.} Training backdoor tasks generally constructs OOD
mappings between backdoor samples and the target class, while keeping the
original in-distribution (ID) mappings between benign samples and the target
class unaffected. When the adversary stops training backdoor tasks, only ID mappings are
introduced and the OOD mappings are gradually erased. In such a case, injecting
subsequent backdoor helps to maintain the previous OOD mappings, thanks to
the shared OOD characteristics relative to the ID data, even for different backdoor triggers and types, and hence the accuracy of the previously planted backdoor can be preserved.

The above observations motivate us to design a novel backdoor detection mechanism which does not rely on statistical comparison of received updates. The idea is that the server in FL can inject certain ``backdoor task'', termed as the indicator task, before broadcasting the global model to the clients. The accuracy of the designed indicator task is expected to quickly fade under benign training at the clients, but will be maintained by backdoors injected at adversarial clients. After receiving all uploaded models, the server can then check the accuracy of the indicator task after correcting the BN
statistics shift. All updates with high accuracy on the indicator task are considered suspicious and filtered out from aggregation. The major challenge here is that the server, in the FL setting, is completely \emph{agnostic} of the unknown backdoors potentially injected by adversarial clients, and we need to design an indicator task, whose accuracy is well maintained by any unknown backdoors. We achieve this goal by proposing a novel backdoor detection
mechanism in FL, BackdoorIndicator, who leverages OOD data to construct the indicator task. We next proceed to describe the details of BackdoorIndicator. 

\subsection{Detailed Methodology}

\begin{algorithm}[!h]
    \caption{BackdoorIndicator}
    \label{algo_bind}
    \KwIn{Indicator Dataset $D_o$, number of training iterations and learning
    rate in training the indicator task: $E$, $\eta$, weight of the
    regularization term $\lambda$ and suspicious threshold $\epsilon_I$. set of
    the selected local client at round $t$: $S_t$.}
    \KwOut{Global model at global round $t+1$: $\boldsymbol{G}^{t+1}$} 
    \tcp{Server initialize at global round $t$}
    The server saves estimated running mean and variance as $\boldsymbol{\mu}_M$
    and $\boldsymbol{\sigma}_M$.
    $\boldsymbol{w}_{ind}\leftarrow \boldsymbol{G}^t$ \\
    \For{$e=1,...,E$}{
        The indicator model 
        $\boldsymbol{w}_{ind}=\boldsymbol{w}_{ind}-\eta\nabla(\mathcal{L}_{task}(\boldsymbol{w}_{ind},D_o)+\lambda||\boldsymbol{w}_{ind}-G^t||_2)$
    }
    The server saves estimated running mean and variance as $\boldsymbol{\mu}_I$
    and $\boldsymbol{\sigma}_I$.

    The server replace the BN statistics in $\boldsymbol{w}_{ind}$ with $\boldsymbol{\mu}_M$
    and $\boldsymbol{\sigma}_M$.

    The server broadcasts $\boldsymbol{w}_{ind}$

    \tcp{Clients perform local training}
    Clients initialize with $\boldsymbol{w}_{ind}$

    \For{client $i\in S_t$}{
        \For{$e_i=1,...,E_i$}{
        Local model
        $\boldsymbol{L}_i=\boldsymbol{L}_i-\eta_i\nabla\mathcal{L}_{task}(\boldsymbol{L}_i,D_b^i)$
        }
    }
    Clients update $\Delta \boldsymbol{w}_i=\boldsymbol{L}_i-\boldsymbol{w}_{ind}$ to the server\\
    \tcp{Server checks indicator accuracy and aggregates}
    $A\leftarrow []$\\
    \For{$i\in S_t$}{       $L_i\leftarrow\boldsymbol{w}_{ind}+\Delta\boldsymbol{w}_i$\\
        $L_i\leftarrow$(replace BN statistics with $\mu_I$ and $\sigma_I$)\\
        $\{\alpha_1,\alpha_2,...,\alpha_N\}\leftarrow$ (check the accuracy of
        the indicator task on $L_i$)\\
        $\alpha_m=max(\{\alpha_1,\alpha_2,...,\alpha_N\})$\\
        \If{$\alpha_m<\epsilon_I$}{
            $A\leftarrow[A;i]$ \tcc*[f]{Accept update $i$}
        }
    }
    $\boldsymbol{G}^{t+1} = \boldsymbol{w}_{ind}+\frac{1}{|A|}\sum_{i\in
    A}\Delta\boldsymbol{w}_i$
\end{algorithm}

To implement BackdoorIndicator, the server first
needs to prepare some OOD data and corresponding indicator task.
In each FL training round, the server starts with saving the BN statistics of the global
model from the last round, denoted as $\boldsymbol{\mu}_M$ and $\boldsymbol{\sigma}_M$. Then the server trains over the OOD data to inject the indicator task, and records the estimated running means $\boldsymbol{\mu}_I$ and variances $\boldsymbol{\sigma}_I$ in BN
layers.
Afterwards, the server replaces the
BN statistics in the trained model with $\boldsymbol{\mu}_M$ and $\boldsymbol{\sigma}_M$
to alleviate any influence of indicator task on the main task performance, and broadcasts the indicator-injected global model to the selected clients.

After the clients finish local
training and upload local models back to the server, the server first replaces
BN statistics in received models with 
$\boldsymbol{\mu}_I$ and $\boldsymbol{\sigma}_I$, and then checks the accuracy
of the indicator task. Any received model with the accuracy exceeding certain
threshold is considered suspicious and will be ruled out from aggregation.
Detailed algorithm is shown in Algorithm 1. We next elaborate details in each
step.

\noindent \textbf{Indicator Task Selection.} The server faces two crucial challenges when designing the indicator task: the server is agnostic of 1) the types and triggers; and 2) the target labels 
of potentially injected backdoors.

To tackle the first challenge, the selected indicator task should be effective
for all potential backdoors, which means that the OOD mapping introduced by the
indicator task should be maintained by any type of injected backdoor. Inspired
by the fact that all backdoor samples are essentially OOD samples with respect to
the benign samples of the target class, we choose a certain amount of
\textit{out-of-distribution} samples, which have distinct real labels with
samples from the benign dataset, to construct the indicator dataset. For
example, we could sample data from EMNIST dataset or from CIFAR100 dataset to construct the indicator dataset for the training on CIFAR10. Specifically, let
$(\boldsymbol{x}_b, y_b)$, $(\boldsymbol{x}_o, y_{ro})$ be the benign and
indicator feature-label pair respectively, where $y_{ro}$ represents the real label of
$\boldsymbol{x}_o$. Also, we denote $\mathcal{Y}_b$, $\mathcal{Y}_{ro}$ as the
benign and indicator label space separately. The server should inject the
indicator task which satisfies $\mathcal{Y}_{b}\cap\mathcal{Y}_{ro}=\emptyset$ to
achieve decent detection performance. Notably, such indicator data is easy to
acquire as it does not require access to the in-distribution data. The server
could adopt public datasets, generated samples or simply randomly generated noise
masks. 

To address the second challenge, while the server does not know the target labels of potential backdoors, it should uniformly sample labels from
the benign label space, and assign the label to each of the indicator samples.
That is, the construction of the indicator dataset $D_0$ is done such that
\begin{equation}
    D_o = \{(\boldsymbol{x}_o^i,y_o^i)\}_{i=1}^N, \quad y_o^i\sim U(\mathcal{Y}_b).
\end{equation}

\noindent \textbf{Injecting Indicator Task.} At the beginning of every FL global round $t$,
the server first saves the estimated running mean $\boldsymbol{\mu}_M$ and
running variance $\boldsymbol{\sigma}_M$ of the global model $G^t$. The server
proceeds to train the indicator task using the constructed indicator dataset, via
optimizing the cross-entropy loss $\mathcal{L}_{task}$. 
To control the influence of injecting the indicator task on the main task, we add a regularization term to punish updates which deviate too much from the
original model. Specifically, let $w_{ind}$ be the
indicator model, the server minimizes the following loss with respect to $D_o$.
\begin{equation}
    \label{eq:inject_loss}
    \mathcal{L} = \mathcal{L}_{task} + \lambda\dot ||w_{ind}-G^t||_2,
\end{equation}
where $\lambda$ denotes the weight of the regularization term. After the server
finishes training the indicator task, it records the estimated running mean
$\boldsymbol{\mu}_I$ and variance $\boldsymbol{\sigma}_I$ of the indicator
model, and then replaces the BN statistics with previously saved
$\boldsymbol{\mu}_M$ and $\boldsymbol{\sigma}_M$. This can help to reduce the influence brought by injecting the indicator task on the main task accuracy, caused by
the difference in the data distribution between the indicator dataset and main task dataset. The server then broadcasts the indicator model to selected
clients, who then conduct local training using their private datasets, and upload the updated local model back to the server.

\noindent \textbf{Inspecting Indicator Accuracy.}  After receiving models from clients, the server proceeds to inspect the performance of the indicator task on each model. As the first step, the server replaces BN statistics of each received model with the
saved $\boldsymbol{\mu}_I$ and $\boldsymbol{\sigma}_I$, correcting the BN statistics shift. Next, the server evaluates the accuracies of samples belonging to different labels in the indicator dataset. Specifically, suppose that the main task has $N = {\cal Y}_b$ classes, the server computes an accuracy array $\boldsymbol{Acc}=\{\alpha_1, \alpha_2,...,\alpha_N\}$, where
$\alpha_i$ represents the test accuracy of all data samples assigned label $i$ in the indicator dataset $D_o$. The server then selects the maximum value
$\alpha_m$ in $\boldsymbol{Acc}$, and regards it as the indicator accuracy. If
the indicator accuracy $\alpha_m$ of a local model exceeds the suspicious
threshold $\epsilon_I$, the model will be marked as backdoor model and
excluded from aggregation. The corresponding label $m$ is considered to be the
backdoor target label of the attacked model. 

\section{Evaluation}
\label{experiments}
In this section, we provide extensive experimental results to demonstrate the
effectiveness of BackdoorIndicator, by comparing its backdoor detection performance with several SOTA backdoor detection mechanisms, including Multi-Krum \cite{blanchard2017machine}, Deepsight
\cite{rieger2022deepsight}, Foolsgold \cite{fung2020limitations}, RFLBAT
\cite{wang2022rflbat} and FLAME \cite{nguyen2022flame}. Specifically, we consider scenarios where an adversary launches backdoor attacks of different types, using different malicious training algorithms, and starting from different global rounds. We also consider the scenarios where the attacker could train the backdoor model using different poisoned learning rates (plrs), and control either a single client or multiple ones in
each global round. We further conduct experiments in FL systems with different non-IID degrees to demonstrate the universal superiority of
BackdoorIndicator. Finally, we reveal how several key hyper-parameters influence
the detection performance of BackdoorIndicator. Our code is available at \url{https://github.com/ybdai7/Backdoor-indicator-defense}.

\subsection{Experiment Setup}
Our experiment implements an FL system running image
classification tasks using FedAVG \cite{mcmahan2017communication}, on a single
machine using an NVIDIA GeForce RTX 4090 GPU with 24GB memory. We evaluate the
performance of the proposed method on three computer vision datasets: CIFAR10,
CIFAR100\cite{krizhevsky2009learning}, and EMNIST\cite{cohen2017emnist} with
three model architecture: VGG16 \cite{simonyan2014very}, ResNet18, and ResNet34\cite{he2016deep}. During the training process, 10 out of total 100
clients are randomly selected to contribute to aggregation in every global
round. For all three datasets, we randomly split the dataset over clients in a
non-IID fashion utilizing Dirichlet sampling\cite{hsu2019measuring}. The
sampling parameter $\alpha$ is set to 0.2 by default to represent a challenging
setting of severe data heterogeneity. We also evaluate BackdoorIndicator under
different non-IID settings.

\noindent \textbf{Evaluation Metric.} We evaluate the detection performance through a set
of metrics: true positive rate (TPR), false positive rate (FPR), and backdoor
accuracy (BA). TPR suggests how well the detection mechanism identifies
adversarial backdoors, which is computed as the ratio of correctly identified
malicious updates to the total number of malicious updates. FPR indicates
how well the detection mechanism discriminates backdoor updates with benign
updates. It is computed as the ratio of benign clients who are mistakenly classified as malicious to the total number of benign updates. As the true negative rate (TNR) and false negative rate (FNR) can be computed as 1 minus FPR and TPR respectively, we only show results of TPR and FPR to evaluate different backdoor defenses methods. BA is
the accuracy of the backdoor task on the global model when the attacker stops poisoning.

\noindent \textbf{Attack Settings.} We validate the defense performance of our proposed
method through considering an attacker who aims to inject \textit{different
types of backdoor} using \textit{different malicious training algorithms}.
Malicious training algorithms and backdoor types are considered as main factors that influence the strength and stealth of the injected backdoor.
\textit{Pixel-pattern backdoors} \cite{gu2020badnets} are the most widely
evaluated backdoor in FL settings, which overlay fixed pixel-patterns over the
original image as the backdoor trigger. The backdoor trigger chosen for
\textit{Blendeded backdoors} \cite{chen2017targeted} is randomly generated noise
map sampled from uniform distribution, which is mixed up with benign images to
construct backdoor samples. While aforementioned backdoors need to modify benign
images to construct backdoor images, \textit{semantic backdoors}
\cite{bagdasaryan01} can be chosen as any naturally occurring feature of the
physical world, and do not require the attacker to modify original images.
Specifically, we choose the semantic backdoor to be the
car-with-vertically-striped-walls-in-the-background \cite{bagdasaryan01} in
CIFAR10. \textit{Edge-case backdoors} \cite{Hongyi01} proceed to poison the
model using data that lives in the tail of the input distribution. Model updates
trained utilizing such data are considered to be less likely to conflict with
other benign updates. This results in a more durable backdoor model against the
vanishing backdoor effect \cite{Zhengming01,chameleon}, and also a more stealthy model which can escape from backdoor detection. 

\begin{table*}[t!]
      \setlength{\abovecaptionskip}{1pt}
      \setlength{\tabcolsep}{4pt}
  \centering
    \caption{Detection performance of all evaluated methods under single client attack with different settings on CIFAR10. The performance is evaluated through the triplet of TPR/FPR(BA). The poisoning lasts for 250 global rounds.}
    \vspace{0.1cm}
    \scalebox{0.75}{
    \begin{tabular}{ccc||ccccccc}
    \toprule
    train alg. & bkdr. types & rds. & No defense & Multi-Krum & Deepsight & Foolsgold & RFLBAT & FLAME & Indicator \\
    \midrule
    \multirow{12}[8]{*}{Vanilla} & \multirow{3}[2]{*}{semantic} & 400   & 0.0/0.0 (46.2) & 0.0/44.3 (71.2) & 4.4/5.7 (38.9) & 49.2/47.9 (18.2) & 0.4/7.3 (33.0) & 0.0/43.5 (44.6) & 72.3/26.0 (0.0) \\
          &       & 800   & 0.0/0.0 (54.3) & 0.0/44.3 (81.4) & 0.8/4.0 (58.1) & 40.4/51.8 (47.7) & 0.4/13.0 (45.4) & 0.0/44.1 (48.6) & 90.1/22.5 (0.0) \\
          &       & 1200  & 0.0/0.0 (83.2) & 0.4/44.2 (81.5) & 3.6/6.5 (60.8) & 54.8/53.4 (60.9) & 1.2/12.5 (63.7) & 0.8/44.1 (95.4) & 86.4/23.9 (0.0) \\
\cmidrule{2-10}    & \multirow{3}[2]{*}{pixel} & 400   & 0.0/0.0 (41.3) & 0.0/44.3 (76.8) & 0.4/2.5 (46.9) & 66.4/48.2 (23.6) & 0.4/6.1 (55.2) & 0.0/44.0 (69.2) & 95.2/25.7 (6.2) \\
          &       & 800   & 0.0/0.0 (71.6) & 0.0/44.2 (80.2) & 8.0/4.6 (46.8) & 68.4/49.9 (47.9) & 0.0/11.6 (68.8) & 0.0/44.1( 87.4) & 95.2/22.2 (9.5) \\
          &       & 1200  & 0.0/0.0 (78.3) & 0.4/44.2 (91.3) & 6.0/6.1 (44.7) & 71.2/51.7 (44.9) & 0.8/13.1 (77.9) & 1.6/44.0 (88.1) & 99.2/15.0 (15.5) \\
\cmidrule{2-10}    & \multirow{3}[2]{*}{blend} & 400   & 0.0/0.0 (46.4) & 0.0/44.3 (65.1) & 3.6/6.2 (56.2) & 3.6/45.4 (64.5) & 0.0/8.3 (79.1) & 0.0/44.0 (70.5) & 63.6/37.2 (29.8) \\
          &       & 800   & 0.0/0.0 (73.3) & 0.0/44.2 (83.8) & 4.0/6.0 (78.7) & 19.6/46.6 (77.3) & 0.0/11.7 (88.3) & 0.4/44.0 (90.2) & 72.4/27.9 (5.1) \\
          &       & 1200  & 0.0/0.0 (79.5) & 0.4/44.2 (90.9) & 8.8/10.0 (72.9) & 28.4/47.7 (74.3) & 0.0/12.3 (90.4) & 2.4/44.0 (92.1) & 83.9/25.0 (7.2) \\
\cmidrule{2-10}    & \multirow{3}[2]{*}{edge} & 400   & 0.0/0.0 (38.12) & 0.0/44.3 (48.1) & 1.2/3.0 (50.0) & 4.0/44.5 (47.7) & 0.0/8.3 (61.2) & 0.0/43.4 (76.6) & 63.6/26.0 (22.5) \\
          &       & 800   & 0.0/0.0 (39.9) & 0.0/44.3 (86.7) & 2.4/5.1 (54.0) & 24.4/48.6 (41.6) & 0.0/11.2 (58.3) & 0.8/44.0 (79.2) & 78.8/18.2 (25.7) \\
          &       & 1200  & 0.0/0/0 (49.4) & 0.0/44.3 (66.5) & 5.2/7.5 (46.2) & 23.6/46.1 (42.4) & 0.4/13.9 (53.8) & 0.0/44.2 (87.8) & 73.2/23.3 (12.1) \\
    \midrule
    \multirow{12}[8]{*}{PGD} & \multirow{3}[2]{*}{semantic} & 400   & 0.0/0.0 (39.2) & 0.0/44.3 (83.6) & 3.2/4.5 (54.7) & 34.4/43.6 (21.4) & 0.4/7.4 (42.6) & 0.0/43.7 (81.4) & 90.8/24.8 (0) \\
          &       & 800   & 0.0/0.0 (43.2) & 0.0/44.3 (72.8) & 7.6/5.8 (12.5) & 26.4/44.4 (53.5) & 0.0/12.3 (60.6) & 0.4/44.0 (97.3) & 73.6/23.5 (0) \\
          &       & 1200  & 0.0/0.0 (60.2) & 0.0/44/3 (86.0) & 3.2/5.0 (38.5) & 38.4/48.5 (44.0) & 0.0/12.0 (57.4) & 0.0/44.3 (96.8) & 74.8/19.6 (0) \\
\cmidrule{2-10}     & \multirow{3}[2]{*}{pixel} & 400   & 0.0/0.0 (48.8) & 0.0/44.3 (73.6) & 8.0/8.6 (31.4) & 66.4/49.1 (10.3) & 0.0/7.6 (52.8) & 0.0/43.8 (73.9) & 77.6/28.0 (23.3) \\
          &       & 800   & 0.0/0.0 (58.2) & 0.0/44.3 (85.5) & 0.2/5.8 (36.3) & 67.6/51.6 (47.0) & 0.0/11.0 (71.6) & 0.0/44.1 (87.3) & 87.6/20.7 (13.5) \\
          &       & 1200  & 0.0/0.0 (71.7) & 0.0/44.3 (92.6) & 8.4/7.9 (45.8) & 71.6/52.6 (56.2) & 0.0/12.4 (79.6) & 0.4/44.0 (94.0) & 94.0/18.9 (12.3) \\
\cmidrule{2-10}    & \multirow{3}[2]{*}{blend} & 400   & 0.0/0.0 (82.1) & 0.0/44.2 (77.3) & 4.0/6.5 (62.3) & 74.0/47.6 (18.2) & 0.0/6.3 (85.2) & 0.0/44.0 (91.3) & 80.8/41.4 (5.2) \\
          &       & 800   & 0.0/0.0 (72.4) & 0.0/44.3 (84.0) & 4.0/4.1 (65.5) & 54.4/49.2 (58.0) & 0.8/10.3 (72.3) & 0.0/44.2 (98.0) & 66.8/25.6 (22.6) \\
          &       & 1200  & 0.0/0.0 (70.1) & 0.0/44.3 (88.4) & 1.6/5.2 (72.3) & 49.2/51.1 (69.4) & 0.4/11.7 (88.3) & 1.2/44.0 (91.9) & 88.8/22.1 (14.7) \\
\cmidrule{2-10}     & \multirow{3}[2]{*}{edge} & 400   & 0.0/0.0 (56.2) & 0.0/44.2 (56.9) & 3.6/5.9 (45.9) & 32.8/47.1 (39.7) & 0.0/7.8 (50.5) & 0.0/43.9 (40.6) & 52.0/24.5 (23.1) \\
          &       & 800   & 0.0/0.0 (60.5) & 0.0/44.3 (84.4) & 2.4/4.5 (60.6) & 52.4/50.3 (44.7) & 0.4/11.6 (45.4) & 0.0/44.1 (68.3) & 64.8/25.6 (13.5) \\
          &       & 1200  & 0.0/0.0 (63.7) & 0.0/44.3 (80.4) & 6.8/8.0 (58.8) & 9.6/47.6 (63.1) & 0.0/14.1 (48.1) & 0.4/44.1 (75.4) & 65.2/20.8 (10.4) \\
    \midrule
    \multirow{12}[8]{*}{Neurotoxin} & \multirow{3}[2]{*}{semantic} & 400   & 0.0/0.0 (74.6) & 0.0/44.2 (85.4) & 4.8/5.7 (13.2) & 29.6/44.0 (31.9) & 0.0/8.1 (26.2) & 0.0/43.2 (70.6) & 71.6/38.2 (0.0) \\
          &       & 800   & 0.0/0.0 (80.2) & 0.0/44.3 (93.1) & 8.0/8.3 (59.0) & 46.4/47.8 (44.0) & 0.4/10.0 (61.9) & 0.8/44.0 (88.8) & 75.6/22.7 (28.2) \\
          &       & 1200  & 0.0/0.0 (93.7) & 0.8/44.2 (91.7) & 8.0/6.8 (56.6) & 58.4/50.5 (62.4) & 0.8/13.1 (74.5) & 0.0/44.2 (90.2) & 72.0/22.0 (27.7) \\
\cmidrule{2-10}     & \multirow{3}[2]{*}{pixel} & 400   & 0.0/0.0 (59.6) & 0.0/44.2 (70.6) & 9.6/8.5 (30.2) & 32.4/48.0 (27.8) & 0.0/6.7 (56.2) & 0.0/43.8 (80.0) & 57.2/38.9 (32.7) \\
          &       & 800   & 0.0/0.0 (74.0) & 0.0/44.3 (80.3) & 6.0/5.4 (34.2) & 39.2/53.1 (64.5) & 0.0/12.0 (69.0) & 0.0/44.1 (87.3) & 95.6/20.2 (5.9) \\
          &       & 1200  & 0.0/0.0 (80.4) & 0.0/44.3 (91.2) & 4.8/5.4 (40.5) & 62.4/45.9 (61.1) & 0.4/11.5 (85.2) & 0.8/44.1 (92.0) & 90.0/20.6 (5.2) \\
\cmidrule{2-10}          & \multirow{3}[2]{*}{blend} & 400   & 0.0/0.0 (78.9) & 0.0/44.2 (89.6) & 4.4/5.6 (55.4) & 42.0/49.1 (29.3) & 0.0/7.6 (63.7) & 0.0/44.0 (73.0) & 44.0/18.3 (26.1) \\
          &       & 800   & 0.0/0.0 (85.3) & 0.0/44.3 (97.4) & 2.4/6.0 (55.5) & 41.2/47.4 (60.1) & 0.0/10.9 (68.6) & 0.0/44.0 (86.2) & 70/17.5 (34.2) \\
          &       & 1200  & 0.0/0.0 (83.6) & 0.0/44.3 (91.4) & 6.0/5.3 (64.3) & 59.2/47.6 (63.3) & 0.4/14.4 (74.5) & 0.8/44.0 (90.6) & 78.4/22.7 (7.4) \\
\cmidrule{2-10}          & \multirow{3}[2]{*}{edge} & 400   & 0.0/0.0 (58.5) & 0.0/44.2 (37.5) & 4.8/4.5 (28.8) & 30.8/43.2 (31.2) & 0.0/6.3 (50.5) & 0.0/43.8 (57.9) & 77.4/28.3 (5.5) \\
          &       & 800   & 0.0/0.0 (46.9) & 0.0/44.3 (58.5) & 5.6/5.8 (47.5) & 37.2/54.2 (46.2) & 0.4/11.8 (54.0) & 0.0/44.0 (69.0) & 50.0/29.7 (25.6) \\
          &       & 1200  & 0.0/0.0 (64.3) & 0.4/44.2 (75.9) & 10.4/7.4 (63.0) & 7.6/47.3 (59.4) & 0.4/15.1 (61.1) & 0.8/44.0 (80.0) & 81.6/17.2 (19.5) \\
    \midrule
    \multirow{12}[8]{*}{Chameleon} & \multirow{3}[2]{*}{semantic} & 400   & 0.0/0.0 (20.6) & 0.0/44.3 (20.7) & 3.2/6.5 (20.8) & 0.0/49.3 (10.2) & 0.0/8.3 (14.8) & 0.0/43.7 (28.6) & 16.8/24.6 (19.6) \\
          &       & 800   & 0.0/0.0 (35.2) & 0.0/44.3 (52.4) & 4.0/4.5 (25.1) & 40.4/45.9 (23.1) & 0.0/12.6 (49.1) & 4.0/43.7 (39.0) & 59.6/23.6 (16.4) \\
          &       & 1200  & 0.0/0.0 (58.7) & 0.0/44.3 (60.9) & 6.4/11.7 (50.0) & 7.6/47.4 (42.0) & 0.0/11.6 (52.9) & 1.6/44.0 (77.1) & 64.0/21.2 (6.2) \\
\cmidrule{2-10}          & \multirow{3}[2]{*}{pixel} & 400   & 0.0/0.0 (58.1) & 0.0/44.3 (61.1) & 9.2/7.4 (34.1) & 51.2/47.8 (32.9) & 0.0/8.4 (43.2) & 1.2/43.7 (66.9) & 74.0/24.6 (19.0) \\
          &       & 800   & 0.0/0.0 (84.6) & 0.4/44.2 (88.4) & 3.2/4.2 (48.8) & 70.0/50.5 (38.1) & 0.4/10.3 (74.5) & 1.2/43.9 (91.1) & 76.4/16.4 (15.1) \\
          &       & 1200  & 0.0/0.0 (80.2) & 0.0/44.2 (95.1) & 4.0/4.0 (55.6) & 57.6/45.9 (58.4) & 0.4/13.1 (83.6) & 5.2/43.5 (90.4) & 81.6/19.4 (18.0) \\
\cmidrule{2-10}          & \multirow{3}[2]{*}{blend} & 400   & 0.0/0.0 (87.0) & 0.0/44.3 (96.9) & 4.4/7.7 (75.0) & 0.8/49.1 (82.0) & 0.0/6.5 (85.5) & 0.0/43.9 (85.7) & 53.2/27.2 (29.7) \\
          &       & 800   & 0.0/0.0 (81.0) & 0.0/44.3 (82.5) & 1.6/5.0 (62.2) & 29.6/50.3 (72.0) & 0.8/10.8 (64.0) & 0.0/44.3 (90.7) & 64.8/24.1 (28.1) \\
          &       & 1200  & 0.0/0.0 (82.3) & 0.4/44.2 (91.2) & 4.8/6.6 (72.5) & 0.0/48.9 (87.0) & 0.0/12.8 (80.2) & 2.8/43.8 (96.1) & 68.8/21.8 (29.8) \\
\cmidrule{2-10}          & \multirow{3}[2]{*}{edge} & 400   & 0.0/0.0 (40.6) & 0.0/44.3 (51.4) & 1.6/5.7 (44.1) & 0.0/45.7 (41.6) & 0.0/8.6 (48.2) & 0.0/44.1 (53.4) & 28.0/27.5 (11.8) \\
          &       & 800   & 0.0/0.0 (53.5) & 0.0/44.3 (61.3) & 8.4/5.2 (43.3) & 48.8/49.2 (44.5) & 0.0/11.4 (50.0) & 0.4/44.1 (93.2) & 56.0/23.9 (19.5) \\
          &       & 1200  & 0.0/0.0 (60.6) & 0.4/44.2 (86.5) & 9.6/9.0 (49.7) & 34.0/46.2 (46.8) & 0.0/11.6 (55.1) & 1.6/44.0 (86.7) & 77.6/18.9 (28.6) \\
    \bottomrule
    \end{tabular}%
    }
  \label{tab:cifar10}%
  \vspace{-0.3cm}
\end{table*}%

As for the malicious training algorithm, an \textit{Vanilla} backdoor attacker
first constructs a malicious training dataset through mixing up backdoor samples
with benign samples; the backdoor model is then
trained on the constructed dataset through optimizing the cross-entropy loss.
\textit{PGD} backdoor attacker, however, trains the backdoor model using
projected gradient descent (PGD) \cite{zitengsun01}, which periodically projects
the model parameters on a ball centered around the model of the previous
iteration, to escape the norm-clipping defense that mitigates the effect from
abnormally large updates. We also consider two malicious training algorithms,
\textit{Neurotoxin} \cite{Zhengming01} and \textit{Chameleon} \cite{chameleon},
which aim to plant more durable backdoors. While Neurotoxin tries to inject the
backdoor using parameters which are not frequently updated by benign clients,
Chameleon enhances the backdoor durability through utilizing sample
relationships, and trains the backdoor model using supervised contrastive
learning \cite{khosla2020supervised}.

We consider scenarios where the backdoor attacker can choose to either corrupt a
single or multiple clients in every global round, and train the backdoor model
with different poisoned learning rates (plr). We additionally evaluate two more
malicious training algorithms, \textit{DBA} \cite{xie2019dba} and \textit{3DFed}
\cite{li20233dfed}, designed especially for multiple client attacks. DBA
decomposes a global pixel-pattern trigger into separated local patterns, and
then embeds them into the dataset of multiple corrupted clients for local
poisoning to achieve a more persistent and stealthy backdoor. 3DFed proposes to
camouflage backdoor models through three well-designed evasion modules, and
dynamically adjust the hyper-parameters through obtaining feedback from previous
global round. 

Also, we do not impose constraints on the global round at which the attacker
initiates poisoning. Specifically, we assume that the attacker may start
poisoning from global rounds 400, 800, or 1200, representing distinct training
stages of the main task. 
Once the poisoning begins, the attacker
consistently participates and injects backdoor models for a specified number of
global rounds, determined according to specific tasks.

\noindent \textbf{Baseline Defenses.} We compare with multiple SOTA backdoor
detection methods described in Section~\ref{sec:defenses}, which are Multi-Krum \cite{blanchard2017machine}, Deepsight
\cite{rieger2022deepsight}, Foolsgold \cite{fung2020limitations}, RFLBAT
\cite{wang2022rflbat} and FLAME \cite{nguyen2022flame}. 
Specifically, as Foolsgold assigns weight for different model updates instead of
directly ruling out suspicious updates, we choose updates whose assigned weights
are smaller than 0.5 to compute the TPR and FPR. 

\noindent \textbf{Indicator Settings.} To successfully implements BackdoorIndicator, the
server needs to sample a certain number of OOD data to construct the indicator
dataset. We randomly select samples from CIFAR100 to construct the indicator
dataset for CIFAR10 task. While for CIFAR100 and EMNIST task, the
indicator dataset is built by samples from CIFAR10. We set the size of the
indicator dataset to 800 by default. For detailed hyper-parameter settings,
the weight of the regulation term $\lambda$ in (\ref{eq:inject_loss}) is set to
0.1, and the suspicious threshold is set to 95 for CIFAR10 and EMNIST
tasks, and 85 for CIFAR100 tasks. At the beginning of every global
round, the server trains the indicator task for 200 iterations. We also vary the
source, size of the indicator dataset, and several key hyper-parameters to reveal
their influence on the detection performance.

\subsection{Detection Performance}
\textbf{Performance under single client attack.} Table \ref{tab:cifar10}
presents the detection performance of all evaluated methods on CIFAR10 under
single client attack with various adversarial settings. Compared with other
detection mechanisms, BackdoorIndicator achieves both the lowest backdoor
accuracy and the highest TPR under all considered adversarial settings. We can also see that the FPR of BackdoorIndicator is only
about the half of the method with the second highest TPR. For vanilla attackers
with the intention to inject pixel-pattern backdoors, BackdoorIndicator
successfully identifies over 95\% of all malicious updates and limits the backdoor accuracy at around 10\%. Specifically, for the attacker who starts
poisoning at 1200 global rounds, BackdoorIndicator achieves 99.2\% TPR and
15.5\% BA, and only misclassifies 15.0\% of benign updates as backdoor updates.
Foolsgold has 44.9\% BA and 71.2\% TPR, which is the second highest of all
evaluated methods. However, to achieve such detection performance, Foolsgold
misclassifies over half of the benign updates as backdoor updates, which is
three times higher than that of BackdoorIndicator. 

\begin{table}[htbp]
      \centering
      \caption{Detection performance of all evaluated methods on CIFAR100 and EMNIST under single client attack. The malicious training algorithm and the backdoor type are Vanilla and pixel-pattern. The adversary starts poisoning from 1200 global round. The poisoning lasts for 250 global rounds.}
      \scalebox{0.95}{
        \begin{tabular}{c||cc}
        \toprule
        methods & CIFAR100 & EMNIST \\
        \midrule
        No defense & 0.0/0.0 (84.5) & 0.0/0.0 (91.8) \\
        Multikrum & 0.0/44.3 (87.8) & 5.0/43.5 (99.1) \\
        Deepsight & 6.4/2.7 (61.3) & 19.0/12.8 (51.4) \\
        Foolsgold & 0.0/0.0 (84.4) & 52.0/46.4 (73.6) \\
        RFLBAT & 0.4/9.7 (85.5) & 1.0/10.0 (90.4) \\
        FLAME & 0.4/44.2 (87.8) & 41.0/39.6 (97.6) \\
        Indicator & \textbf{85.2}/15.8 \textbf{(7.3)} & \textbf{99.0}/35.4 \textbf{(9.6)} \\
        \bottomrule
        \end{tabular}%
        }
        \label{tab:cifar100emnist}%
\end{table}%

\begin{table*}[!h]
      \setlength{\abovecaptionskip}{0pt}
      \centering
      \caption{Detection performance of all evaluated methods on CIFAR10 under single client attack with different non-IID settings and poisoned learning rate (plr). The malicious training algorithm and the backdoor type are Vanilla and pixel-pattern. The adversary starts poisoning from 1200 global round. The performance is evaluated through the triplet of TPR/FPR (BA). The poisoning lasts for 250 global rounds.}
    \vspace{0.1cm}
    \scalebox{0.8}{
    \begin{tabular}{cc||ccccccc}
    \toprule
    alpha & plr & No defense & Multi-Krum & Deepsight & Foolsgold & RFLBAT & FLAME & Indicator \\
    \midrule
    \multirow{4}[2]{*}{0.2} & 0.01  & 0.0/0.0 (66.8) & 0.0/44.2 (86.7) & 10.4/6.0 (31.3) & 76.8/53.1 (24.8) & 0.4/12.2 (67.3) & 0.0/44.2 (83.0) & 98.0/19.7 (11.5) \\
          & 0.025 & 0.0/0.0 (78.3) & 0.4/44.2 (91.3) & 6.0/6.1 (44.7) & 71.2/51.7 (44.9) & 0.8/13.1 (77.9) & 1.6/44.0 (88.1) & 99.2/15.0 (15.5) \\
          & 0.04  & 0.0/0.0 (79.3) & 7.6/43.4 (93.5) & 7.6/7.1 (46.2) & 60.0/55.4 (74.7) & 1.6/12.9 (90.3) & 38.8/40.1 (90.8) & 96.0/21.2 (18.6) \\
          & 0.055 & 0.0/0.0 (88.1) & 100.0/33.6 (9.3) & 7.6/7.4 (49.8) & 47.6/47.8 (84.0) & 2.0/13.1 (88.0) & 100.0/33.6 (13.0) & 97.6/21.2 (9.6) \\
    \midrule
    \multirow{4}[2]{*}{0.5} & 0.01  & 0.0/0.0 (74.6) & 0.0/44.3 (87.4) & 19.2/6.3 (40.1) & 54.8/39.6 (68.5) & 1.6/14.5 (69.4) & 0.0/44.3 (87.7) & 99.2/9.9 (10.7) \\
          & 0.025 & 0.0/0.0 (88.3) & 0.0/44.3 (89.9) & 7.6/5.4 (68.0) & 65.6/36.5 (79.2) & 0.0/13.4 (75.8) & 0.8/44.2 (88.9) & 89.6/13.1 (23.3) \\
          & 0.04  & 0.0/0.0 (90.0) & 0.4/44.2 (94.5) & 11.6/8.5 (60.9) & 29.2/48.0 (90.4) & 0.0/4.7 (89.1) & 1.6/44.2 (93.0) & 87.6/19.5 (34.1) \\
          & 0.055 & 0.0/0.0 (92.5) & 5.2/43.7 (93.9) & 10.8/6.5 (71.2) & 47.2/41.3 (88.1) & 1.2/13.6 (91.2) & 99.6/33.7 (10.5) & 87.2/21.2 (26.3) \\
    \midrule
    \multirow{4}[2]{*}{0.9} & 0.01  & 0.0/0.0 (84.5) & 0.0/44.3 (87.5) & 23.2/10.6 (68.2) & 56.8/34.5 (59.1) & 0.0/11.6 (85.6) & 0.0/44.3 (89.3) & 100.0/9.5 (9.6) \\
          & 0.025 & 0.0/0.0 (84.2) & 0.0/44.3 (89.7) & 22.8/10.8 (76.3) & 43.6/33.1 (84.5) & 1.6/13.7 (87.5) & 0.4/44.2 (90.2) & 92.4/6.6 (13.3) \\
          & 0.04  & 0.0/0.0 (90.1) & 2.0/44.0 (92.5) & 16.4/6.8 (78.7) & 39.2/29.3 (89.0) & 0.0/10.7 (89.7) & 1.2/44.1 (91.6) & 90.0/19.4 (47.6) \\
          & 0.055 & 0.0/0.0 (90.1) & 24.4/41.7 (93.7) & 11.6/5.3 (77.1) & 43.6/33.1 (90.0) & 1.6/12.0 (90.6) & 54.4/38.4 (90.6) & 89.6/18.6 (56.2) \\
    \bottomrule
    \end{tabular}%
    }
    \label{tab:alphas_plrs}%
    \vspace{-0.4cm}
\end{table*}%

BackdoorIndicator can effectively detect more stealthy backdoors injected
by vanilla adversaries. Specifically, BackdoorIndicator could identify blended
backdoors with 83.9\% TPR and 7.2\% BA, and edge-case backdoors with 73.2\% TPR
and 12.1\% BA. However, Foolsgold can only successfully detect 28.4\% and 23.6\%
of backdoor updates for blended backdoor and edge-case backdoor respectively,
resulting 74.3\% and 42.4\% BA for these two backdoors. 

BackdoorIndicator can also successfully identify backdoor updates trained with more advanced algorithms. For PGD attackers who inject pixel-pattern backdoor
from 1200 global rounds, BackdoorIndicator filters out 94.0\% of all backdoor updates to achieve 12.3\% BA, while the largest TPR of the rest methods is 71.6\% with 56.2\% BA. BackdoorIndicator achieves the best performance for pixel-pattern backdoor updates trained using Neurotoxin and Chameleon from 1200 global rounds, with the TPR/FPR (BA) triplet as 90.0/20.6 (5.2) and 81.6/19.4 (18.0) respectively. 

BackdoorIndicator can identify backdoor planted from different training stages.
Vanilla attackers can successfully plant blended backdoors against other methods
through poisoning model starting from 400 global rounds, which is the early
training stage. The highest TPR and the lowest BA for all evaluated methods is
3.6\% and 56.2\%, compared with BackdoorIndicator which detects 63.6\% of backdoor
updates and achieves 29.8\% BA. Attackers could also adopt more advanced
algorithm to increase BA for blended backdoor. Chameleon attackers
achieve a BA higher than 75\% against other
methods. However, they still fail to bypass BackdoorIndicator, which achieve
53.2\% TPR and 29.7\% BA. These results not only demonstrate the strong capability of BackdoorIndicator in identifying backdoors, but also suggests that it can effectively discriminate between backdoor and benign updates. This could help to protect the global model from backdoor poisoning, while maintaining a fast and accurate training of the main task.

BackdoorIndicator can still effectively reduce the
backdoor accuracy, even when its detection performance declines
as adversaries inject backdoors in the early training stage.
For blended backdoor updates trained using Neurotoxin, the TPR of BackdoorIndicator drops from 70\% to 44.0\% if the attacker starts poisoning
from 400 global rounds rather than 800. However, BackdoorIndicator
still achieves a lower BA of 26.1\% with 44.0\% TPR. We argue that it is due to the difficulty of injecting backdoors into a model that is far from convergence: in the following training process, benign updates are more likely to conflict with backdoor updates, resulting in a slow increase of BA even when there is no defense. 

Table \ref{tab:cifar10} also reveals that, BackdoorIndicator's ability of
precisely distinguishing backdoor updates and benign updates gets stronger if
the attack happens later in the training stage. To defend a vanilla pixel-pattern backdoor attack, the
FPR of BackdoorIndicator decreases from 25.7\% to 22.2\%, and further to 15.0\% as the attack starts later. This is because benign updates becomes smaller in magnitude and closer to each other in distance, resulting in a lower false alarm rate.

\begin{table*}[htbp]
      \setlength{\abovecaptionskip}{0pt}
  \centering
  \caption{Detection performance of all evaluated methods on CIFAR10 under multiple client attack. The backdoor type are Vanilla and pixel-pattern. The adversary starts poisoning from global round 1200 for CIFAR100, and 400 for EMNIST. The performance is evaluated through the triplet of TPR/FPR (BA). The poisoning ends until 360 backdoor models are injected.}
    \vspace{0.1cm}
    \scalebox{0.8}{
    \begin{tabular}{cc||ccccccc}
    \toprule
    train alg.& bkdr. \% & No defense & Deepsight & Foolsgold & RFLBAT & RFLBAT & FLAME & Indicator \\
    \midrule
    \multirow{3}[2]{*}{Vanilla} & 20\%  & 0.0/0.0 (88.4) & 11.7/9.7 (56.6) & 99.7/42.5 (8.5) & 0.0/16.8 (85.5) & 0.0/16.8 (85.5) & 0.0/49.0 (94.5) & 97.2/23.5 (10.6) \\
          & 40\%  & 0.0/0.0 (89.2) & 5.6/17.2 (71.7) & 100.0/36.1 (9.5) & 0.1/19.8 (88.6) & 0.1/19.8 (88.6) & 0.0/61.6 (97.3) & 95.8/14.8 (14.7) \\
          & 60\%  & 0.0/0.0 (91.3) & 10.0/44.4 (95.3) & 100.0/30.3 (9.5) & 0.1/34.4 (91.9) & 0.1/34.4 (91.9) & 0.0/77.8 (97.4) & 93.8/19.4 (6.1) \\
    \midrule
    DBA   & 40\%  & 0.0/0.0 (87.1) & 21.1/15.8 (85.2) & 99.7/35.8 (7.8) & 0.0/22.3 (75.2) & 0.0/22.3 (75.2) & 0.0/62.3 (94.4) & 98.6/21.1 (4.8) \\
    \midrule
    3DFed & 20\%  & 0.0/0.0 (78.6) & 16.7/10.0 (36.8) & 99.1/42.1 (9.2) & 0.0/14.8 (70.8) & 0.0/14.8 (70.8) & 3.3/48.4 (93.0) & 98.3/20.3 (6.7) \\
    \bottomrule
    \end{tabular}%
    }
  \label{tab:multiple_attack}%
  \vspace{-0.2cm}
\end{table*}%

We find that all evaluated methods except for BackdoorIndicator and
Foolsgold fails to detect backdoor updates. We argue that it is because of the
challenging adversarial training setting and non-IID degree of the data. It is hard for these methods to effectively detect backdoor models trained using a small
learning rate in highly non-IID settings. We follow up to study their influence on the detection performance.

An interesting observation in Table \ref{tab:cifar10} is that the utilization of alternative backdoor detection techniques, such as FLAME, would unexpectedly lead to an improvement in backdoor accuracy. This is because the considered attackers could fabricate poisoned models to make them statistically close to the benign ones by using small learning rate. In such scenarios, statistical methods, like FLAME, fail to detect any backdoor updates, while still mark and filter out around 44\% benign updates. This will lead to a decrease in the number of accepted updates, and consequently an increase in the aggregation weights of backdoor updates, which is computed as one divided the total number of accepted updates. This eventually leads to an improvement in backdoor accuracy compared with scenarios without any defenses.

We also compare the detection performance on more datasets. We present a part of
results in Table \ref{tab:cifar100emnist} and the full results in
Appendix \ref{appendix}. The proposed method still achieves the lowest BA, and the highest TPR
for both CIFAR100 and EMNIST. Specifically, BackdoorIndicator identifies 85.2\%
of backdoor updates, which restricts backdoor accuracy at 7.3\%, and only
misclassifies 15.8\% benign models as malicious on CIFAR100. However, other
evaluated methods fail to filter out backdoor updates, among which Deepsight
achieves the highest TPR as 6.4\% and the lowest BA as 61.3\%. The performance
of these methods improves on EMNIST with the highest TPR as 52.0\% and the
lowest BA as 51.4\%. Nevertheless, BackdoorIndicator still achieves the best
performance with the TPR/FPR (BA) triplet as 99.0\%/35.4\% (9.6\%). This further
demonstrates the universal effectiveness of ackdoorIndicator across different
datasets.

\begin{table}[htbp]
      \setlength{\abovecaptionskip}{0pt}
      \centering
      \caption{Performance on different model architectures under single client attack for CIFAR10 task. The backdoor type are Vanilla and pixel-pattern. The adversary starts poisoning from 1200 global round, and continues for 250 rounds.}
      \vspace{0.1cm}
      \scalebox{0.95}{
        \begin{tabular}{c||cc}
        \toprule
        architecture & ResNet34 & VGG16 \\
        \midrule
        No defense & 0.0/0.0 (77.3) & 0.0/0.0 (76.7) \\
        Multikrum & 0.7/44.1 (88.4) & 0.7/44.1 (86.6) \\
        Deepsight & 11.3/6.5 (42.2) & 6.7/4.3 (56.4) \\
        Foolsgold & 61.3/46.8 (31.7) & 21.3/35.0 (76.2) \\
        RFLBAT & 0.0/14.0 (61.4) & 0.7/12.3 (79.2) \\
        FLAME & 0.0/44.0 (91.4) & 1.3/43.9 (85.2) \\
        Indicator & \textbf{91.9}/38.0 \textbf{(14.3)} & \textbf{95.3}/10.5 \textbf{(13.6)} \\
        \bottomrule
        \end{tabular}%
        }
      \label{tab:diff_architect}%
      \vspace{-0.4cm}
    \end{table}%

\noindent \textbf{Performance under different non-IID degrees and poisoned
learning rates.} Table \ref{tab:alphas_plrs} exhibits detection performance of
all evaluated methods under different non-IID degrees and multiple poisoned
learning rates (plr). BackdoorIndicator can stably identify malicious updates
regardless of the change in plr. For the highly non-IID setting of $\alpha
=0.2$, BackdoorIndicator detects 98.0\% backdoor updates when plr equals 0.01,
and also achieve 97.6\% TPR with 0.055 plr. However, the change in plr
substantially influences the performance of the other methods. FLAME and
Multi-Krum can effectively identify backdoor updates trained using large plr
while fail to detect any malicious update trained using small plr. This is
because that these two methods identify potential malicious updates based on
certain distance measure, and updates with abnormal distance features will be
marked as outliers. In such a case, it is much harder for these methods to
detect backdoor models trained using small plr which makes the backdoor
models closer to the global model. In contrast, Foolsgold and Deepsight show
weaker detection ability as the plr increases. Specifically for Foolsgold, as it assigns lower weights for updates that have strong similarity with others,
backdoor updates trained using larger plr exhibit larger differences from benign
updates, making them harder for Foolsgold to identify.

We also note that BackdoorIndicator consistently performs well under different
non-IID settings, with an average TPR of over 90\%. For the rest methods,
Foolsgold exhibits weaker performance in more IID setting, while Deepsight shows
stronger sensitivity in identifying backdoor updates when $\alpha$ grows larger.
FLAME shows better performance as non-IID degree gets stronger. Under the
setting where plr equals 0.055, FLAME achieves 100\% TPR when $\alpha$ equals
0.2 and 54.5\% TPR when $\alpha$ equals 0.9.

\noindent \textbf{Performance under multiple client attack.} Table
\ref{tab:multiple_attack} presents the detection performance when attackers
control multiple malicious clients in every global round. We find that the
proposed method can successfully defend against backdoor attacks even when 60\% of
the clients 
are malicious. BackdoorIndicator achieves
over 93\% TPR and limits BA at around 10\% in all the evaluated tasks. We also
note that Foolsgold can identify over 99\% of backdoor updates in all evaluated
multiple client attack settings, which shows stronger detection ability compared with when facing single client attack. However, the FPRs of Foolsgold
are still about twice compared with the FPR of BackdoorIndicator. This  
indicates that BackdoorIndicator can precisely discriminate between benign and
backdoor updates even under multiple client attack.

\noindent \textbf{Performance for different model architectures.} We also conduct experiments
to demonstrate the effectiveness of BackdoorIndicator on different model architectures. As is shown in Table \ref{tab:diff_architect}, the detection
performance of BackdoorIndicator remains the strongest for both ResNet34 and VGG16, yielding the lowest BA among all evaluated methods. 

\begin{table}[htbp]
  \centering
  \vspace{-0.4cm}
  \caption{Main task accuracy \textbf{(LEFT)} when applying BackdoorIndicator, and \textbf{(RIGHT)} when not applying any defense mechanism, under single client attack with different malicious training algorithms, backdoor types, and injection rounds on CIFAR10. The poisoning lasts for 250 global rounds.}
  \vspace{0.1cm}
  \scalebox{0.75}{
    \begin{tabular}{cc||ccc}
    \toprule
    Train alg. & bkdr types & 400   & 800   & 1200 \\
    \midrule
    \multirow{4}[2]{*}{Vanilla} & semantic & 82.3/82.1 & 86.2/86.4 & 87.0/88.7 \\
          & pixel & 82.2/80.5 & 82.6/85.5 & 87.2/87.9 \\
          & blend & 81.9/81.1 & 84.0/83.6 & 86.9/87.4 \\
          & edge  & 82.4/79.4 & 85.2/84.9 & 87.2/88.2 \\
    \midrule
    \multirow{4}[2]{*}{PGD} & semantic & 80.3/82.5 & 86.4/85.1 & 87.6/86.9 \\
          & pixel & 79.4/80.3 & 85.8/86.2 & 87.2/87.0 \\
          & blend & 81.0/82.8 & 82.7/85.8 & 86.6/87.3 \\
          & edge  & 82.9/80.1 & 86.5/87.9 & 86.1/86.5 \\
    \midrule
    \multirow{4}[2]{*}{Neurotoxin} & semantic & 79.8/81.3 & 82.3/83.7 & 86.7/86.0 \\
          & pixel & 80.9/79.3 & 85.2/85.6 & 87.1/87.3 \\
          & blend & 79.5/82.9 & 84.6/84.1 & 86.1/86.9 \\
          & edge  & 81.9/81.0 & 83.6/82.5 & 86.3/88.1 \\
    \midrule
    \multirow{4}[2]{*}{Chameleon} & semantic & 81.5/78.6 & 83.8/84.3 & 87.3/86.4 \\
          & pixel & 79.2/79.9 & 86.3/85.6 & 87.2/87.9 \\
          & blend & 78.8/80.8 & 85.3/87.3 & 86.8/86.0 \\
          & edge  & 80.7/83.4 & 85.4/86.0 & 87.3/86.9 \\
    \bottomrule
    \end{tabular}
    }
    \vspace{-0.4cm}
  \label{tab:mainacc}%
\end{table}%

\noindent \textbf{BackdoorIndicator does not degrade main task accuracy.} As it is shown in Table \ref{tab:mainacc}, we further provide the main task accuracy with the same setting in Table \ref{tab:cifar10}, when applying BackdoorIndicator and when not applying any defense mechanism. We can see that although the proposed defense slightly impacts the main task accuracy in some cases, applying BackdoorIndicator achieves almost identical main task accuracy for most settings to when not applying any defenses.

\subsection{Impact of Hyper-parameters}
We proceed to reveal the influence of several key hyper-parameters on the performance of
BackdoorIndicator. In the following experiments, we assume that vanilla
attackers start to inject backdoors from 1200 global rounds, and keep poisoning
for 100 global rounds. We repeat each task for 5 times, and present the mean value and the standard deviation.
\begin{table}[htbp]
      \setlength{\abovecaptionskip}{0pt}
      \centering
      \caption{Performance of BackdoorIndicator which collects data from different sources to construct the indicator dataset. MA* indicates the main task accuracy without replacing BN statistics with the main BN statistics after injecting the indicator task. The size of the indicator dataset is 800.}
      \vspace{0.05cm}
      \scalebox{0.85}{
        \begin{tabular}{c||ccc}
        \toprule
        source & TPR/FPR & MA* & MA \\
        \midrule
        CIFAR100 & 95.6$\pm$3.7/22.4$\pm$3.6 & 84.3$\pm$2.4 & 84.8$\pm$1.4 \\
        EMNIST & 93.0$\pm$4.2/32.3$\pm$1.7 & 72.7$\pm$3.1 & 87.0$\pm$1.3 \\
        300k random & 95.8$\pm$1.7/19.0$\pm$1.4 & 84.1$\pm$3.0 & 83.1$\pm$1.0 \\
        random noise & 90.0$\pm$7.0/41.3$\pm$8.6 & 69.6$\pm$4.2 & 85.2$\pm$1.7 \\
        \bottomrule
        \end{tabular}%
        }
      \label{tab:source}%
      \vspace{-0.2cm}
    \end{table}%
    
\noindent \textbf{Influence of the source of indicator dataset.} Table
\ref{tab:source} provides the performance of the proposed method which collects
data from different sources to build the indicator dataset. We find that \emph{the detection performance is rather insensitive to the data source that is used to construct the indicator dataset}. Constructing the indicator dataset using data
from EMNIST and even uniformly sampled random noise can achieve 93.0\% TPR and 90.0\% TPR
respectively. However, the FPR in these scenarios, which are 32.3\% and 41.3\%,
increases substantially compared to 22.4\% FPR and 19.0\% FPR, in which case the
defender samples data from CIFAR100 and 300k random images
\cite{hendrycks2022pixmix} to build the indicator dataset respectively. 

We also note that BN statistics is crucial in correctly evaluating the main task
accuracy (MA). 
For the indicator dataset which consists of data sampled from EMNIST and random noise, the MA, without replacing BN statistics with the main BN statistics after injecting the indicator task, is around 72.7\% and 69.6\% respectively; while
the MAs are about 87.0\% and 85.2\% when using main BN statistics. On the other hand, the
influence of whether replacing BN statistics with the main BN statistics is relatively negligible for indicator dataset with CIFAR100 and 300K random data. Utilizing the BN statistics from the indicator dataset may cause degradation of the main task, especially for an indicator dataset that is considerably different from the main dataset.
Thus, it is crucial for the defender, who is agnostic to the data distribution of the main task, to replace BN statistics with the main BN statistics to avoid potential drop of MA.

\begin{table}[htbp]
      \setlength{\abovecaptionskip}{0pt}
      \centering
      \caption{Performance of BackdoorIndicator with different sizes of the indicator dataset.}
      \vspace{0.05cm}
      \scalebox{0.9}{
        \begin{tabular}{c||c}
        \toprule
        size & TPR/FPR \\
        \midrule
        100   & 96.6$\pm$4.8/53.2$\pm$4.2 \\
        200   & 88.6$\pm$10.7/23.5$\pm$2.3 \\
        500   & 93.2$\pm$6.8/23.7$\pm$2.7 \\
        800   & 95.6$\pm$3.7/22.4$\pm$3.6 \\
        1100  & 96.2$\pm$3.6/19.3$\pm$2.3 \\
        \bottomrule
        \end{tabular}%
        }
      \label{tab:size}%
      \vspace{-0.4cm}
    \end{table}

\noindent \textbf{Influence of the size of indicator dataset.} We provide experiments
to explore how the size of the indicator dataset influences the detection
performance. As is shown in Table \ref{tab:size}, the decrease in the size of the
indicator dataset causes negligible influence in identifying the backdoor updates. With merely 100 samples in the indicator dataset, BackdoorIndicator is able to 
identify around 96.6\% of all malicious updates. Nevertheless, the FPR with 100 samples is around 53.2\%, indicating a weak ability in precisely discriminating between benign and malicious
updates.
This problem can be effectively addressed through having 200 samples in
the indicator dataset. In such a case, BackdoorIndicator achieves around 23.5\%
FPR, which is about the same performance when the size equals 800. This suggests that the defender does not need to collect a large number of indicator data to implement BackdoorIndicator. 

For main tasks with more classes, the defender generally needs to construct a relatively large indicator dataset to ensure sufficient number of samples for all potential adversary-chosen labels. 
However, evaluation results in Table \ref{tab:cifar100emnist} show that implementing BackdoorIndicator with the indicator dataset of 800 samples can secure the CIFAR100 main task from backdoor attacks with the TPR/FPR (BA) triplet as 85.2/15.8 (7.3). The detection performance with approximately 8 OOD-samples-per-class for CIFAR100 is comparable to applying the indicator dataset with 200 samples in Table \ref{tab:size} for CIFAR10, which corresponds to 20 OOD-samples-per-class. This is because the indicator task is much easier to be forgotten by the benign updates, when there are more classes. Therefore, to achieve comparable difference between the indicator task accuracies of the benign and backdoor updates (and hence similar backdoor detection performance), the task with a dataset of more classes can afford to use a smaller number of OOD samples per class. 
This also suggests that the size of the indicator dataset does not need to increase linearly with the number of classes, further demonstrating the practicality of BackdoorIndicator.

\begin{table}[htbp]
      \setlength{\abovecaptionskip}{0pt}
      \centering
      \caption{Performance of BackdoorIndicator with different numbers of training iterations.}
      \vspace{0.05cm}
      \scalebox{0.9}{
        \begin{tabular}{c||c}
        \toprule
        training iterations & TPR/FPR \\
        \midrule
        10    & 95.0$\pm$1.4/16.0$\pm$2.8 \\
        200   & 95.6$\pm$3.7/22.4$\pm$3.6 \\
        \bottomrule
        \end{tabular}%
        }
      \label{tab:iterations}%
      \vspace{-0.3cm}
    \end{table}%

\noindent \textbf{Influence of indicator training iterations.} We also provide experiments
to show that implementing the proposed method does not add too much computation
overhead to the FL system. Table \ref{tab:iterations} shows that training
the indicator task for 10 iterations can achieve around 95.0\% TPR and 16.0\%
FPR, which is comparable with training for 200 iterations. 

The above set of experiments demonstrates that the indicator dataset is easy to construct: no strict requirements on the data source (even random noise suffices), and only a few hundred samples are needed; and the training of indicator task is lightweight: only a few iterations are needed. These features make the deployment of BackdoorIndicator widely feasible in practical FL systems. 

\begin{table}[htbp]
      \setlength{\abovecaptionskip}{0pt}
      \centering
      \caption{Performance of BackdoorIndicator with different values of $\lambda$.}
      \vspace{0.1cm}
      \scalebox{0.9}{
        \begin{tabular}{c||cc}
        \toprule
        $\lambda$ & TPR/FPR & MA \\
        \midrule
        0     & 89.0$\pm$3.7/20.7$\pm$5.7 & 76.1$\pm$8.2 \\
        0.01  & 96.3$\pm$2.3/19.3$\pm$4.8 & 81.0$\pm$3.4 \\
        0.1   & 95.6$\pm$3.7/22.4$\pm$3.6 & 84.8$\pm$1.4 \\ 
        0.2   & 94.7$\pm$3.3/18.8$\pm$7.8 & 86.8$\pm$0.4 \\
        0.3   & 97.0$\pm$1.6/14.9$\pm$3.6 & 86.0$\pm$2.2 \\
        \bottomrule
        \end{tabular}%
        }
      \label{tab:lambda}%
      \vspace{-0.2cm}
    \end{table}%

\noindent \textbf{Influence of $\lambda$ in training the indicator task.} Table \ref{tab:lambda}
illustrates the performance of the proposed method with different $\lambda$, which
is the weight of the regularization term in training the indicator task. We
observe a sizable decrease in MA as $\lambda$ decreases.
Specifically, BackdoorIndicator achieves around 76.1\% MA when we exclude the
regularization term from training. The MA gradually increases to 84.8\% and
further to 86.8\% as $\lambda$ is increased to 0.1 and 0.2 respectively. Without
proper regularization in the change of the model parameter, the model gradually
forgets the main task, causing a drop in the MA. Thus, a sufficiently large
$\lambda$ needs to be chosen to make BackdoorIndicator effective. 

\noindent \textbf{Influence of the suspicious threshold $\epsilon_I$}. A
proper suspicious threshold needs to be chosen to achieve a decent performance
in both identifying backdoor updates and discriminating with benign updates. For
the same task, decreasing $\epsilon_I$ undoubtedly boosts TPR and FPR
simultaneously, resulting in a stronger ability in detecting malicious updates while a weaker ability in distinguishing benign updates. For different
tasks, as is shown in Table \ref{tab:cifar10} and
\ref{tab:cifar100emnist}, BackdoorIndicator achieves comparable performance for
CIFAR100 with $\epsilon_I=85$ and CIFAR10 with $\epsilon_I=95$ under the same adversarial setting. This is because that the maintaining effect of clients' backdoors on the indicator task is weaker when there are more classes, which causes the accuracy of the indicator task to decline and a relatively smaller $\epsilon_I$ is needed. 
\vspace{-0.3cm}
\section{Resilience to Adaptive Attacks}
\vspace{-0.2cm}
Knowing how BackdoorIndicator operates, adversaries may adopt adaptive attacks to bypass the detection. We consider such attack
settings and strategies utilizing \textit{pre-training}.

\noindent \textbf{Pre-training.} As shown in Table \ref{tab:alphas_plrs}, the
increase in the poisoned learning rates results in the decrease of the detection
performance of BackdoorIndicator. This is
because the indicator task is forgotten more quickly when the attacker trains a backdoor model with a larger learning rate. Thus, adaptive attackers could pre-train the model using its benign dataset to first erase the indicator task, and then train the model using the constructed backdoor
dataset to escape from the detection of BackdoorIndicator.

We proceed to demonstrate the performance of the above adaptive attack against
BackdoorIndicator. We assume that an adaptive attacker first trains the model
using the benign dataset for 10 iterations with 0.05 learning rate. The attacker
then trains the backdoor model using the constructed malicious dataset with a
plr of 0.025. We assume that the attacker adopts a vanilla training algorithm and
intends to inject pixel-pattern backdoors. We repeat each task for 3 times, and
report the mean value and the standard deviation. The evaluated dataset is
CIFAR10. 

\begin{table}[htbp]
      \setlength{\abovecaptionskip}{0pt}
  \centering
  \caption{Performance of BackdoorIndicator w./wo. norm clipping defense (NCD) against adaptive attack.}
      \vspace{0.05cm}
    \scalebox{0.9}{
    \begin{tabular}{c||c}
    \toprule
    method & TPR/FPR (BA) \\
    \midrule
    Indicator wo. NCD& 53.0$\pm$9.6/19.3$\pm$4.2 (39.1$\pm$15.1) \\
    Indicator w. NCD & 75.3$\pm$8.0/13.6$\pm$8.2 (9.9$\pm$1.8) \\
    \bottomrule
    \end{tabular}%
    }
  \label{tab:adaptive}%
  \vspace{-0.25cm}
\end{table}%

As is shown in Table \ref{tab:adaptive}, BackdoorIndicator only identify 53.0\%
of all malicious updates and achieves around 39.1\% BA, indicating the
effectiveness of the adaptive attack. To this end, we further modify the
proposed defense to improve resilience to the above adaptive attack.
Specifically, for the pre-training step to be effective, the uploaded local model of an adaptive attack needs to be sufficiently different from the global model with indicator task injected. Motivated by this, we propose
to leverage norm clipping defense (NCD), which clips the norm of each received
update to a fixed bound, to mitigate the forgetting of the indicator task, potentially introduced by adaptive attackers through pre-training.  
We observe from Table~\ref{tab:adaptive} that BackdoorIndicator equipped with NCD can now effectively detect around 75\% of backdoor updates, and limit the BA to merely around 9.9\%. 

\section{Discussion}
\label{discussion}
The proposed BackdoorIndicator can be composed with other statistical defenses for FL, further improving effectiveness. For example, BackdoorIndicator can replace the dynamic
clustering component in FLAME to serve as a more precise backdoor detector, which could help to further minimize the amount of noise needed in FLAME. Given that the design principle of BackdoorIndicator based on detection of OOD task injection is orthogonal to those of statistical methods, incorporating BackdoorIndicator is expected to provide multiplicative gains on defense performance. Nevertheless, while it is not the focus of this paper, we envision that future works
could seek to combine BackdoorIndicator with other detection methods to build stronger FL backdoor defense mechanisms.

Despite the effectiveness of  BackdoorIndicator corroborated by empirical evaluations, it is also important to understand the theoretical explanation behind the maintaining effect on injected backdoors. We envision that future research could provide theoretical analysis to further facilitate the understanding of BackdoorIndicator, guiding the development of stronger backdoor detection mechanisms.

\section{Related Work}
\label{relatedwork}
\textbf{Federated Learning.} Multiple variants of FedAVG are proposed to tackle
the data heterogeneity problem in FL. FedProx \cite{li2020federated} argues that
simply adding a proximal term to constrain local updates from deviating
from the global model could help to enhance the performance under non-IID
setting. SCAFFOLD \cite{karimireddy2020scaffold} tries to mitigate the
statistical heterogeneity through correcting client drifts using control
variates. Another line of work, named personalized FL, attempts to handle the
heterogeneity through training distinct models for different clients. Per-FedAVG \cite{per_fedavg} utilize
the Model-Agnostic Meta-Learning (MAML) framework to find an initial global model
that local clients can easily adapt to their dataset through performing a few
steps of gradient descent. pFedMe \cite{t2020personalized} formulates the problem in a bi-level
way, and solve it using Moreau envelops as clients' loss function, which
achieves better performance compared with Per-FedAVG.\\
\textbf{Backdoor attack} is an emerging security threat in training deep neural
networks (DNNs). Except for the BadNets \cite{gu2020badnets}, which first reveal
the vulnerability of DNNs against backdoor attack, a series of works proposes
more stealthy backdoor attacks to escape from detection algorithms. The
adversarial backdoor embedding algorithm \cite{shokri2020bypassing} not only
optimizes the original loss function of the model, it also seeks to maximize the
indistinguishability of the hidden representations of poisoned data and clean
data. Composite attacks \cite{lin2020composite} proposes to use backdoor
triggers composed from existing benign features of multiple labels to elude
detection. Li \textit{et al.}\cite{li2021invisible} explores sample-specific
backdoor attack, where the backdoor trigger for each sample is distinct. The
method only needs to modify a small amount of perturbation to original images,
which makes the trigger invisible. Apart from the successful implementation of
backdoor attacks in poisoning DNNs, it has also been shown to effectively attack
other deep learning paradigms. In continual learning setting, Kang \textit{et
al.} \cite{kang2023poisoning} shows that attacker could plant an input-aware
backdoor to stealthily promote forgetting on the previous task while retaining high
accuracy at the current task. Yang \textit{et al.} \cite{yang2023data} shows
that multi-modal models are also vulnerable to poisoning attacks.\\ 
\textbf{Backdoor defenses} in training deep learning neural networks can be
mainly categorized into: pruning-based defenses, trigger synthesis based
defenses, and saliency map based defenses. For pruning based methods, Liu et al.
\cite{liu2018fine} proposes to prune backdoor-related neurons to remove
backdoors based on the observation that these neurons are usually dormant during
the inference process. Neural cleanse \cite{wang2019neural} is the first
proposed trigger synthesis based method, which first obtains potential backdoor
trigger for each class, and proceeds to reconstruct the final trigger and its
target label through anomaly detection. The following work
\cite{qiao2019defending} argues acquiring the entire trigger distribution is
essential for effective defending against backdoor attacks. Thus, the method
adopts generative modeling to recover the whole trigger distribution. For
saliency map based defenses, SentiNet \cite{chou2020sentinet} utilizes the
Grad-CAM \cite{selvaraju2017grad} to extract critical regions from input, and
then locate the trigger regions based on the boundary analysis. NeuronInspect
\cite{huang2019neuroninspect} exhibits a similar idea through identifying the
existence of backdoor attacks by generating heatmaps from the output layer. The
method proceeds to extract key features and apply an anomaly detector to
identify backdoor updates.\\
\noindent \textbf{Backdoor attacks and defenses in FL}. We further review several recent works on backdoor attacks and defenses in FL, which are not evaluated in this paper. F3BA \cite{Fang_Chen_2023} tries to mount backdoor attacks through only compromising a small fraction of least important parameters, and further optimize the trigger to improve effectiveness. CerP \cite{Lyu_Han_Wang_Liu_Wang_Liu_Zhang_2023} casts the distributed backdoor attack as a joint optimization process to reduce the bias between benign and backdoor updates. These two works try to fabricate malicious updates to make them statistically close to benign ones to escape from backdoor detection, which utilize the inherent weakness of statistical backdoor defenses as revealed in our paper. However, the proposed backdoor samples are still OOD samples with respect to benign ones from the target class. Planting these backdoors will still maintain the indicator accuracy, and eventually be identified and filtered out by BackdoorIndicator. 

A recent work \cite{krauss2024automatic} proposes a novel backdoor attack method, AutoAdapt, which leverages the Augmented Lagrangian method for constraint optimization to create backdoor models which can automatically adapt to various defense metrics to evade detection. However, AutoAdapt is not applicable in attacking BackdoorIndicator. This is because the indicator dataset is kept secret in the server, and is agnostic to the attacker, which renders the attacker impossible from computing the defense metric to adapt backdoor models to evade the BackdoorIndicator detection.

For other backdoor defense mechanisms in FL, BayBFed \cite{10179362} relies on a Bayesian statistical method to filter out backdoor updates, which still suffers from the inherent weakness of statistical backdoor defenses. ADFL \cite{zhu2023adfl} utilizes GAN to revise the global model and eliminate the planted backdoors. However, this method requires the server to access a part of clean training data, which is rather impractical in FL. 

\vspace{-0.2cm}
\section{Conclusion}
\label{conclusion}

We propose a novel proactive backdoor detection method BackdoorIndicator, to
detect potential backdoors injected to the clients' updates in federated
learning. BackdoorIndicator builds upon the maintaining effect of subsequent
backdoors on the previous ones and the fact that backdoor samples are OOD
samples compared to benign samples from the target class, 
designs indicator task leveraging OOD samples to identify and rule out backdoor
updates. Extensive experiments show the superior performance and practicality of
BackdoorIndicator, under a wide range of system and adversarial settings.

\section*{Acknowledgement}
This work is in part supported by the National Nature Science Foundation of China (NSFC) Grant 62106057.

\bibliographystyle{plain}
\bibliography{backdoorindicator}

\appendix
\section*{Appendix}

\section{Supplementary Experiments}
\label{appendix}

We provide supplementary experimental results to show the effectiveness of
BackdoorIndicator under various adversarial and system settings on CIFAR100 and
EMNIST. As it is shown in Table \ref{tab:cifar100_1}, \ref{tab:emnist_1},
\ref{tab:cifar100_2}, \ref{tab:emnist_2}, BackdoorIndicator consistently achieves the best performance across all evaluated settings.

\begin{table*}[!h]\footnotesize
      \setlength{\tabcolsep}{5pt}
    \centering
    \caption{Detection performance of all evaluated methods under single client attack with different settings on CIFAR100. The performance is evaluated through the triplet of TPR/FPR(BA). The poisoning lasts for 100 global rounds.}
      \begin{tabular}{cc||ccccccc}
      \toprule
      train alg. & bkdr types & No defense & Multi-Krum & Deepsight & Foolsgold & RFLBAT & FLAME & Indicator \\
      \midrule
      \multirow{3}[2]{*}{Vanilla} & pixel & 0.0/0.0 (84.5) & 0.0/44.3 (87.8) & 6.4/2.7 (61.3) & 0.0/0.0 (84.4) & 0.4/9.7 (85.5) & 0.4/44.2 (87.8) & 85.0/15.8 (7.3) \\
            & blend & 0.0/0.0 (57.8) & 0.0/44.0 (68.1) & 10.0/4.1 (29.3) & 0.0/0.0 (50.6) & 1.0/11.4 (50.0) & 0.0/44.0 (68.7) & 80.0/20.5 (1.7) \\
            & edge  & 0.0/0.0 (73.3) & 0.0/44.0 (81.8) & 7.0/3.7 (70.7) & 0.0/0.0 (80.8) & 0.0/9.2 (78.5) & 0.0/44.0 (71.6) & 77.0/19.3 (10.4) \\
      \midrule
      \multirow{3}[2]{*}{PGD} & pixel & 0.0/0.0 (74.4) & 0.0/44.0 (80.0) & 9.0/4.0 (31.7) & 0.0/0.0 (77.5) & 0.0/11.2 (73.7) & 0.0/44.0 (73.8) & 100.0/20.1 (0.5) \\
            & blend & 0.0/0.0 (64.8) & 0.0/44.0 (58.5) & 13.0/5.9 (21.2) & 0.0/0.0 (62.1) & 0.0/11.6 (54.9) & 0.0/44.0 (62.3) & 100.0/21.6 (0.3) \\
            & edge  & 0.0/0.0 (75.0) & 0.0/44.0 (88.7) & 7.0/2.0 (60.6) & 0.0/0.0 (67.8) & 0.0/8.4 (81.4) & 0.0/44.0 (84.9) & 100.0/18.9 (0.0) \\
      \midrule
      \multirow{3}[2]{*}{Neurotoxin} & pixel & 0.0/0.0 (73.0) & 0.0/44.0 (81.4) & 5.0/2.2 (50.9) & 0.0/0.0 (75.7) & 1.0/10.8 (77.8) & 0.0/44.0 (79.1) & 99.0/20.4 (1.4) \\
            & blend & 0.0/0.0 (72.3) & 0.0/44.0 (77.6) & 9.0/2.9 (21.7) & 0.0/0.0 (57.6) & 0.0/10.4 (64.7) & 0.0/44.0 (77.5) & 89.0/19.6 (0.9) \\
            & edge  & 0.0/0.0 (72.5) & 0.0/44.0 (81.1) & 7.0/3.8 (69.8) & 0.0/0.0 (69.9) & 0.0/9.6 (64.7) & 0.0/44.0 (89.6) & 85.0/18.6 (3.1) \\
      \midrule
      \multirow{3}[2]{*}{Chameleon} & pixel & 0.0/0.0 (61.2) & 0.0/44.0 (72.4) & 10.0/4.5 (35.1) & 0.0/0.0 (57.3) & 1.0/9.7 (64.6) & 0.0/44.0 (64.6) & 100.0/20.9 (0.9) \\
            & blend & 0.0/0.0 (12.1) & 0.0/44.0 (33.3) & 4.0/0.8 (21.1) & 0.0/0.0 (12.8) & 1.0/11.5 (14.7) & 0.0/44.0 (22.0) & 79.0/23.9 (1.5) \\
            & edge  & 0.0/0.0 (30.7) & 0.0/44.0 (61.4) & 13.0/5.7 (34.2) & 0.0/0.0 (20.4) & 0.0/9.3 (47.4) & 0.0/44.0 (50.8) & 93.0/21.3 (0.0) \\
      \bottomrule
      \end{tabular}%
    \label{tab:cifar100_1}%
  \end{table*}%

\begin{table*}[!h]\footnotesize
    \centering
    \caption{Detection performance of all evaluated methods under single client attack with different settings on EMNIST. The performance is evaluated through the triplet of TPR/FPR(BA). The poisoning lasts for 100 global rounds.}
      \begin{tabular}{c||ccccccc}
      \toprule
      train alg. & No defense & Multi-Krum & Deepsight & Foolsgold & RFLBAT & FLAME & Indicator \\
      \midrule
      Vanilla & 0.0/0.0 (91.8) & 5.0/43.5 (99.1) & 19.0/12.8 (51.4) & 52.0/46.4 (73.6) & 1.0/10.0 (90.4) & 41.0/39.6 (97.6) & 99.0/35.4 (9.6) \\
      PGD   & 0.0/0.0 (96.3) & 4.0/43.6 (99.9) & 15.0/10.3 (40.4) & 56.0/50.6 (62.3) & 0.0/11.1 (99.1) & 20.0/41.7 (99.7) & 92.0/32.4 (10.1) \\
      Neurotoxin & 0.0/0.0 (81.8) & 8.0/43.2 (99.7) & 18.0/11.3 (57.1) & 39.0/51.2 (97.4) & 1.0/8.5 (82.6) & 17.0/42.1 (96.2) & 100.0/36.8 (10.1) \\
      Chameleon & 0.0/0.0 (92.5) & 5.0/43.5 (97.6) & 6.0/10.1 (99.9) & 48.0/49.0 (82.3) & 1.0/11.9 (99.8) & 19.0/41.9 (9.7) & 98.0/35.1 (10.1) \\
      \bottomrule
      \end{tabular}%
    \label{tab:emnist_1}%
  \end{table*}%

\begin{table*}[!h]\footnotesize
    \centering
    \caption{Detection performance of all evaluated methods on CIFAR100 under single client attack with different non-IID settings and poisoned learning rate (plr). The malicious training algorithm and the backdoor type are Vanilla and pixel-pattern. The adversary starts poisoning from 1200 global round. The performance is evaluated through the triplet of TPR/FPR (BA). The poisoning lasts for 100 global rounds.}
      \begin{tabular}{cc||ccccccc}
      \toprule
      alpha & plr & No defense & Multikrum & Deepsight & Foolsgold & RFLBAT & FLAME & Indicator \\
      \midrule
      \multirow{3}[2]{*}{0.2} & 0.01  & 0.0/0.0 (62.9) & 0.0/44.0 (73.3) & 8.0/2.3 (38.5) & 0.0/0.0 (65.1) & 2.0/12.0 (61.8) & 0.0/44.0 (63.1) & 100.0/18.1 (0.6) \\
            & 0.03  & 0.0/0.0 (78.7) & 3.0/43.7 (80.1) & 9.0/2.2 (34.5) & 0.0/0.0 (75.5) & 1.0/12.7 (78.1) & 0.0/44.0 (76.3) & 94.0/20.7 (1.2) \\
            & 0.05  & 0.0/0.0 (82.7) & 82.0/35.8 (50.6) & 6.0/2.2 (38.6) & 0.0/0.0 (83.7) & 4.0/8.4 (82.8) & 100.0/34.0 (0.9) & 85.0/22.0 (13.9) \\
      \midrule
      \multirow{3}[2]{*}{0.5} & 0.01  & 0.0/0.0 (63.7) & 0.0/44.0 (74.9) & 6.0/2.1 (34.2) & 0.0/0.0 (61.4) & 0.0/10.7 (66.2) & 0.0/44.0 (60.8) & 99.0/16.1 (0.7) \\
            & 0.03  & 0.0/0.0 (78.4) & 2.0/43.8 (83.5) & 0.0/0.0 (42.4) & 0.0/0.0 (75.4) & 2.0/8.4 (80.9) & 1.0/43.9 (78.4) & 96.0/12.2 (0.9) \\
            & 0.05  & 0.0/0.0 (83.5) & 100.0/34.0 (0.9) & 7.0/1.3 (32.2) & 0.0/0.0 (79.5) & 3.0/11.3 (83.3) & 100.0/34.0 (0.9) & 88.0/12.9 (6.3) \\
      \midrule
      \multirow{3}[2]{*}{0.9} & 0.01  & 0.0/0.0 (68.5) & 0.0/44.0 (72.8) & 3.0/0.4 (38.4) & 0.0/0.0 (66.0) & 1.0/10.0 (66.7) & 0.0/44.0 (55.4) & 100.0/12.1 (0.5) \\
            & 0.03  & 0.0/0.0 (78.9) & 5.0/43.5 (81.6) & 6.0/2.4 (40.0) & 0.0/0.0 (78.1) & 0.0/9.0 (78.3) & 1.0/43.9 (72.6) & 97.0/10.6 (1.2) \\
            & 0.05  & 0.0/0.0 (81.2) & 100.0/34.0 (0.9) & 4.0/1.4 (44.7) & 0.0/0.0 (82.1) & 5.0/9.8 (83.5) & 100.0/34.0 (0.9) & 87.0/12.2 (8.4) \\
      \bottomrule
      \end{tabular}%
    \label{tab:cifar100_2}%
  \end{table*}%

\begin{table*}[!h]\footnotesize
    \centering
    \caption{Detection performance of all evaluated methods on EMNIST under single client attack with different non-IID settings and poisoned learning rate (plr). The malicious training algorithm and the backdoor type are Vanilla and pixel-pattern. The adversary starts poisoning from 1200 global round. The performance is evaluated through the triplet of TPR/FPR (BA). The poisoning lasts for 100 global rounds.}
      \begin{tabular}{cc||ccccccc}
      \toprule
      alpha & plr & No defense & Multikrum & Deepsight & Foolsgold & RFLBAT & FLAME & Indicator \\
      \midrule
      \multirow{3}[2]{*}{0.2} & 0.01  & 0.0/0.0 (73.1) & 0.0/44.0 (97.4) & 11.0/7.0 (42.8) & 44.0/47.9 (79.2) & 0.0/12.5 (63.6) & 4.0/43.5 (99.6) & 100.0/36.7 (10.1) \\
            & 0.03  & 0.0/0.0 (82.1) & 12.0/42.8 (96.1) & 16.0/17.0 (68.8) & 47.0/51.4 (94.7) & 0.0/10.8 (97.2) & 89.0/35.0 (10.4) & 98.0/40.5 (10.0) \\
            & 0.05  & 0.0/0.0 (93.1) & 73.0/36.7 (99.8) & 24.0/18.3 (69.0) & 51.0/52.0 (55.7) & 3.0/12.3 (95.8) & 100.0/33.8 (10.1) & 97.0/41.5 (10.1) \\
      \midrule
      \multirow{3}[2]{*}{0.5} & 0.01  & 0.0/0.0 (77.8) & 1.0/43.9 (99.5) & 22.0/9.1 (97.5) & 34.0/42.7 (91.1) & 0.0/8.4 (91.2) & 8.0/43.0 (99.8) & 100.0/28.4 (10.1) \\
            & 0.03  & 0.0/0.0 (98.9) & 21.0/41.9 (99.9) & 31.0/18.1 (94.2) & 38.0/37.5 (99.5) & 0.0/9.9 (99.5) & 99.0/34.1 (10.3) & 99.0/25.2 (10.0) \\
            & 0.05  & 0.0/0.0 (99.1) & 100.0/34.0 (10.0) & 23.0/14.2 (99.8) & 31.0/42.3 (99.6) & 4.0/10.7 (98.9) & 100.0/34.0 (10.0) & 95.0/27.7 (10.0) \\
      \midrule
      \multirow{3}[2]{*}{0.9} & 0.01  & 0.0/0.0 (96.0) & 1.0/43.9 (97.7) & 14.0/5.1 (83.4) & 43.0/32.5 (93.6) & 0.0/10.8 (96.8) & 10.0/43.0 (99.9) & 100.0/21.3 (10.0) \\
            & 0.03  & 0.0/0.0 (95.6) & 38.0/40.2 (99.8) & 28.0/15.9 (85.2) & 49.0/35.3 (96.7) & 0.0/11.9 (99.9) & 96.0/34.0 (10.0) & 100.0/26.9 (10.0) \\
            & 0.05  & 0.0/0.0 (99.9) & 98.0/34.2 (10.0) & 19.0/12.7 (98.4) & 55.0/32.9 (99.5) & 3.0/11.0 (99.7) & 100.0/34.0 (9.9) & 100.0/21.7 (10.0) \\
      \bottomrule
      \end{tabular}%
    \label{tab:emnist_2}%
  \end{table*}%

\end{document}